\documentclass[a4paper,fleqn,usenatbib]{mnras}

\usepackage{newtxtext,newtxmath}

\usepackage[T1]{fontenc}
\usepackage{ae,aecompl}

\usepackage{graphicx}	
\usepackage{amsmath}	
\usepackage{amssymb}	
\usepackage{tikz}

\title[External disc photoevaporation in 2D]{The first multi-dimensional view of mass loss from externally FUV irradiated protoplanetary discs }

\author[Haworth \& Clarke]
{\parbox{\textwidth}{Thomas J. Haworth$^{1}$\thanks{E-mail: \texttt{t.haworth@imperial.ac.uk}} and  Cathie J. Clarke$^2$ 
}\vspace{0.4cm}\\
\parbox{\textwidth}{$^{1}$ Astrophysics Group, Imperial College London, Blackett Laboratory, Prince
Consort Road, London SW7 2AZ, UK \\
$^{2}$ Institute of Astronomy, Madingley Rd, Cambridge, CB3 0HA, UK \\
}}

\begin{document}

\date{Accepted ???. Received ???; in original form ???}

\pagerange{\pageref{firstpage}--\pageref{lastpage}} \pubyear{2016}

\maketitle
\label{firstpage}

\begin{abstract}
Computing the flow from externally FUV irradiated protoplanetary discs requires solving complicated and expensive photodissociation physics iteratively in conjunction with hydrodynamics. Previous studies have therefore been limited to 1D models of this process. In this paper we compare 2D-axisymmetric {models of externally photoevaporating discs} with their 1D analogues, finding that mass loss rates are consistent to within a factor four. The mass loss rates in 2D are higher, in part because half of the mass loss comes from the disc surface (which 1D models neglect). 1D mass loss rates used as the basis for disc viscous evolutionary calculations are hence expected to be conservative. We study the anatomy of externally driven winds including the streamline morphology, kinematic, thermal and chemical structure. A key difference between the 1D and 2D models is in the chemical abundances. For instance in the 2D models CO can be dissociated at smaller radial distances from the disc outer edge than in 1D calculations because gas is photodissociated by radiation along trajectories that are assumed infinitely optically thick in 1D models. Multidimensional models will hence be critical for predicting observable signatures of environmentally photoevaporating protoplanetary discs.

\end{abstract}


\section{introduction}
The detection of almost 4000 very diverse exoplanets, most of which were discovered in the last 5 years \citep[e.g.][]{2015ARA&A..53..409W, 2018haex.bookE.195W}, coupled with new observing facilities such as the Atacama Large Millimetre Array (ALMA) has motivated and made possible a revolution in our understanding of planet formation. There is now overwhelming evidence that planets form from discs of material around young stars \citep[for reviews see e.g.][]{2011ARA&A..49...67W, 2015PASP..127..961A, 2016PASA...33....5O, 2016PASA...33...59S, 2016PASA...33...53H, 2017RSOS....470114E}. The vast majority of studies of disc evolution and planet formation consider the young star-disc in isolation and such approaches are proving very effective at understanding the myriad physical processes that take place in planet-forming discs.

The young disc-hosting stars themselves typically form in clustered groups of hundreds to hundreds of thousands \citep[e.g.][]{2010RSPTA.368..713L, 2018arXiv181201615K}. In order to fully understand the planet formation process, and in particular the populations of resulting planets, we also have to understand how the natal cluster can affect discs and any planet formation within them. Specifically, any given planet-forming disc can be influenced either by gravitational encounters or irradiation by other cluster members \citep[for a recent assessment of their relative impacts, see][]{2018MNRAS.478.2700W}. For example \cite{2018MNRAS.474..886N} accounted for stellar neighbours through simple passive heating of the outer disc in planetary population synthesis models, finding that this is necessary to suppress large populations of cold Jupiters (particularly at low metallicity) which are not observed.

In particular, UV irradiation by stellar neighbours can heat a disc and drive material away from it in a photoevaporative wind.  If this mass loss is higher than the rate of viscous spreading, the disc will also be truncated \citep[e.g.][]{2007MNRAS.376.1350C, 2013ApJ...774....9A, 2017MNRAS.468L.108H, 2017MNRAS.468.1631R, 2018MNRAS.478.2700W, 2018ApJ...867...41S}. Although the effect of external photoevaporation is expected to act most directly upon the outer disc, limiting the disc mass reservoir and radial extent affects the ability of material to move into the inner disc which may also have consequences for planet formation at small orbital radii \citep[e.g. as in the case of Trappist-1][]{2017A&A...604A...1O, 2018MNRAS.475.5460H}. Furthermore, truncation of the disc lowers the viscous timescale and affects the redistribution of angular momentum and hence surface density evolution of \textit{the entire disc}, an effect which does not occur through internal photoevaporation by the host star, where instead an inner hole is eventually produced \citep{2001MNRAS.328..485C, 2010MNRAS.401.1415O}. The effect of external photoevaproation on planet formation should therefore not be discounted simply because it operates on the disc at larger radii than those at which most planets are being discovered.

The range of star forming  environemts in the Galaxy implies that stars may be exposed to ambient ultraviolet fluxes that vary over five orders of magnitude \citep[e.g.][]{2008ApJ...675.1361F}; it is therefore possible that some of the exoplanet diversity being discovered is influenced by diversity in the properties of the natal cluster \citep[e.g.][]{2019arXiv190211094N}. Note that although gravitational interactions in the cluster do take place, generally, the effect of external UV irradiation has a dominant influence \citep[][]{2001MNRAS.325..449S, 2016arXiv160501773G, 2018MNRAS.478.2700W}. Internal winds are also driven by the host star \citep[see e.g. the review by][]{2017RSOS....470114E} though the relative impact and interplay between internally and externally driven winds is currently unknown.

A major complication is that environmental evaporation of discs is only easy to observe and/or model in extreme UV conditions, for example in the vicinity of O stars. In such a regime disc evaporation is unsubtle, giving rise to ``proplyds'' that can be observed in the optical, silhouetted against the H\,\textsc{ii} region \citep[e.g.][]{1996AJ....111.1977M, 1998AJ....115..263O, 2001AJ....122.2662O, 2002ApJ...566..315H, 2012ApJ...746L..21W}. Trends in disc properties such as mass and/or radius near such sources are also starting to be inferred \citep[e.g.][]{2017AJ....153..240A, 2018ApJ...860...77E}. 

The majority of star/discs are not in such an extreme UV environment as in the vicinity of an O star. For example \cite{2008ApJ...675.1361F} compute probability distributions for the UV environments of Galactic star forming regions. Proplyds lie in the tail end of this distribution, even in the limit of no extinction (and hence more pervasive high UV fields). In the weak-intermediate UV regime environmental disc photoevaporation is more subtle, making it difficult to observe. 

It is also much more difficult to model environmental photoevaporation in more modest UV environments, with the thermal state of the gas being sensitive to the far-ultraviolet (FUV) radiation field and photodissociation region (PDR) chemistry \citep[e.g.][]{2004ApJ...611..360A}. In particular, a key issue is that the main coolant is the escape of line photons, which cannot be assumed to be optically thin. To estimate this the escape probability into three dimensions has to be computed from every point on a computational domain \citep[e.g.][]{2012MNRAS.427.2100B}. Although within the disc itself this can be approximated by assuming some dominant trajectory \citep[e.g. vertically and/or outward radially][]{2017ApJ...847...11W, 2018ApJ...857...57N, 2018arXiv181012330W} in a photoevaporative wind this is not necessarily the case. A 3D estimate is required, which makes multidimensional models difficult to develop and, historically, computationally prohibitive. Given that there is  growing observational evidence for external disc photoevaporation in more modest environments with the detection of proplyds around a B star in Orion \citep{2016ApJ...826L..15K} and a possible photoevaporative halo around IM Lup \citep{2016ApJ...832..110C, 2017MNRAS.468L.108H, 2018A&A...609A..47P} further models and observations in these lower UV regimes are essential. 

The difficulty of including 3D line cooling means that modelling external disc photoevaporation in the FUV regime has been restricted to 1D models to date \citep[though at very high UV field strengths the role of the FUV can be more easily approximated in 2D axisymmetric calculations][]{2000ApJ...539..258R}. These assume that mass loss occurs predominantly from the disc outer edge \citep[see e.g.][ for a discussion on this]{2004ApJ...611..360A}, since this part of the disc is a substantial mass reservoir that is least gravitationally bound to the star. The exciting UV radiation is assumed to propagate from the outside inwards and vice versa for the line cooling. 
Initially these models were computed semi-analytically, locating the critical point of the flow and matching to conditions
at the flow base (disc outer edge), with the temperature calculated from a pre-computed table as a function of local density, UV field and extinction \citep{2004ApJ...611..360A, 2011PASP..123...14H, 2016MNRAS.457.3593F}. Although solutions are not always possible with this approach, where they are possible mass loss rates  that would be significant for disc evolution are predicted (though all such models only considered Solar type stars). 

In recent years direct simulations of evaporating discs have become possible \citep{2016MNRAS.463.3616H, 2017MNRAS.468L.108H, 2018MNRAS.475.5460H, 2018MNRAS.481..452H}. These have still been 1D, but directly compute the PDR chemical/thermal structure and hydrodynamics. For example \cite{2018MNRAS.481..452H} produced a publicly available grid of steady state mass loss rates for a large variety of stellar, disc and UV parameters. A key general result is that mass loss can be important even in very weak UV environments if the disc is only weakly bound, for example in the case of a very extended disc like IM Lup \citep{2017MNRAS.468L.108H} or if the stellar mass is low as would have been the case for the precursor of a Trappist-1 like system \citep{2018MNRAS.475.5460H}. Although these 1D models are easily calculated with modern facilities, they offer only a weak means of predicting the crucial signatures of externally evaporating discs which we require in order to identify the process in action in a range of UV environments. Multidimensional models are required to effectively predict observables. Furthermore, since the 1D models do predict significant mass loss, even in weak UV environments, they must be validated and improved upon with more realistic multidimensional models. 

In this paper we present the first 2D cylindrical models of externally FUV irradiated protoplanetary discs. There is a large parameter space where the nature of external photoevaporation may differ (e.g. as a function of UV field strength, stellar mass, disc mass, disc size, metallicity, and so on) so in this paper we focus on only a handful of first calculations. We aim to gain a first insight into the applicability of 1D mass loss rates, as well as test the validity of assumptions in the 1D models such as the mass predominantly being driven from the disc outer edge. We also aim to study the flow morphology and take a first look at the chemical composition of 2D externally driven winds.

\section{Numerical method}
We begin with an overview of the methodology that permits us to solve PDR-dynamics in arbitrary geometries with 3D line cooling. 

\subsection{General Overiew}
We use the photochemical-hydrodynamics code \textsc{torus-3dpdr} \cite{2015MNRAS.454.2828B} to run the models in this paper. In its most general form this is a coupling of the grid-based Monte Carlo radiative transfer and hydrodynamics code \textsc{torus} \citep{2000MNRAS.315..722H, 2012MNRAS.420..562H, 2015MNRAS.453.2277H, 2015MNRAS.448.3156H, 2018MNRAS.477.5422A} with the 3D photodissociation region modelling code \textsc{3d-pdr} \citep{2012MNRAS.427.2100B}\footnote{Though note that the Monte Carlo radiation transport is not actually employed within this paper, as we will discuss in section  \ref{sec:pdrChemTherm}.}. In this hybridisation all calculations take place on the \textsc{torus} grid. That is, components have been stripped from \textsc{3d-pdr} and directly incorporated into  \textsc{torus}.   The original idea behind the coupled code is that the Monte Carlo radiation transport of \textsc{torus} can accurately compute the exciting UV radiation field (which \textsc{3d-pdr} approximates as, e.g. isotropic, planar or spherical) for the PDR modules, which is important for models of the interstellar medium. However the application of using the PDR thermal structure in hydrodynamical problems such as external disc photoevaporation has been a powerful by-product that has dominated the use of \textsc{torus-3dpdr} to date. 

PDR-dynamics calculations are performed by iteratively computing hydrodynamics and PDR equilibirium updates (i.e. via operator splitting), which we now summarise separately. 

\subsection{Hydrodynamics}
The hydrodynamics scheme is a finite volume method which employs a \cite{vanleer} flux limiter, \cite{1983AIAAJ..21.1525R} interpolation and is total variation diminishing. In these calculations the disc masses are low enough that self-gravity is unimportant, so we include only a point source gravitational potential. The hydrodynamics scheme was first presented in \textsc{torus} and detailed in \cite{2012MNRAS.420..562H}. 

\subsection{Equilibrium PDR chemistry and thermal balance}
\label{sec:pdrChemTherm}
The PDR {aspect of our} calculations require solving a chemical network iteratively with thermal balance \citep[for full details see][]{2012MNRAS.427.2100B, 2015MNRAS.454.2828B}. The fact that the PDR calculations in \textsc{3d-pdr} take place in 3D for arbitrary geometries is crucial for our models of externally photoevaporating protoplanetary discs. To compute the line cooling escape probabilities, $4\,\pi$ steradians is sampled from the centre of every cell on the computational domain using a \textsc{healpix} scheme \citep{2005ApJ...622..759G} which divides the sky into zones of equal solid angle. The size of these zones is dependent on the level of \textsc{healpix} refinement $l$, with a number of zones $N_l=12\times 4^l$. The optical depth, escape probability, etc. along each of these rays is then computed to estimate the line cooling. These rays are also used to estimate the local UV, ``$\chi$'',  (i.e. we are not using the Monte Carlo radiative transfer of \textsc{torus}) as attenuated from an isotropic ambient UV field $\chi_0$ according to 
\begin{equation}
	\left(\frac{\chi}{1\,G_0}\right) = \frac{1}{N_l}\sum_1^{N_l}\chi_0\exp(-\tau_{UV}).
\end{equation}
where $\tau_{UV}$ is the UV optical depth, which is related to the visual extinction by $\tau_{UV}=3.02{A_V}$. {$G_0$ is the Habing unit of UV field strength \cite{1968BAN....19..421H}.} {Although our models are 2D-axisymmetric, \textsc{healpix} sampling of the sky is still done in 3D utilising the symmetry of the problem. }

In its current form our \textsc{healpix} scheme uses a long characterstics ray tracing, for which the scaling is relatively poor. Multiple rays have to be traced from every single cell on the grid, which is computationally expensive. Furthermore, since there can be significant memory requirements we implement domain decomposition (the grid memory is distributed amongst cores) which means rays may not have access to the information stored on all parts of the grid along their trajectory. Most domains hence have to act as servers (sending relevant information to the domain doing the ray tracing) and take turns doing ray-tracing, rather than tracing rays simultaneously. This is added to the fact that the PDR calculation takes place over many iterations to converge on the equilibrium chemical and thermal structure, so the ray tracing process has to take place many times for any given single PDR update. Of course this then has to be done multiple times iteratively with hydrodynamics updates. The 3D line cooling/PDR calculation therefore overwhelmingly dominates the computational expense of these models.

Given the poor scaling we generally run these calculations on a small number of cores. For example on a $256^2$ cell 2D cylindrical grid (where the ray tracing takes place in 3D, utilising the symmetry of the problem) a single PDR update at the lowest level of \textsc{healpix} refinement can have a wall time of around 12-24 hours on 5 MPI threads (with remaining threads on the node being used for shared memory parallelisation).  In future we intend to improve on the wall time per PDR calculation by improving the scaling to larger numbers of cores. In light of the above, we are presently limited to calculating the steady state flow structure from an irradiated disc, and also at modest resolution both spatially and in terms of \textsc{healpix} rays. Nevertheless in this paper we do consider convergence with both of these quantities in section \ref{sec:convergence}. 

\begin{table}
 \centering
  \caption{A summary of the species included and initial gas abundances {\citep[taken from][]{2009ARA&A..47..481A} } for the reduced network used in this paper, which consists of 33 species and 330 reactions \citep{2012MNRAS.427.2100B}. The sum of  hydrogen atoms in atomic and molecular hydrogen is unity. The other abundances are with respect to atomic hydrogen.}
  \label{PDRGuts}
  \begin{tabular}{@{}l c l c@{}}
   \hline
   \hline    
  {Gas} \\
  \hline
   Species & Initial abundance & Species & Initial abundance \\
   \hline
   H & $4\times10^{-1}$ & H$_2$ & $3\times10^{-1}$ \\
   He & $8.5\times10^{-2}$ & C+ & $2.692\times10^{-4}$\\
   O & $4.898\times10^{-4}$ & Mg+ & $3.981\times10^{-5}$ \\
   H+ & 0 & H$_2$+ & 0 \\
   H$_3$+ & 0& He+ & 0 \\
   O+ & 0 & O$_2$ & 0 \\
   O$_2$+ & 0 & OH+ & 0 \\
   C & 0 & CO & 0 \\
   CO+ & 0 & OH & 0 \\
   HCO+& 0 & Mg & 0 \\
   H$_2$O& 0 & H$_2$O+ & 0 \\
   H$_3$O & 0 & CH & 0 \\
   CH+ & 0 & CH$_2$ & 0 \\
   CH$_2$+ & 0 & CH$_3$ & 0 \\
   CH$_3$+ & 0 &CH$_4$ & 0 \\ 
   CH$_4$+ & 0 & CH$_5$+ & 0 \\    
   e$^-$  & 0 \\      
   \\      
   \hline    
   \hline    
   {Dust} \\
   \hline
    $\sigma_{\textrm{FUV}}$ & $2.7\times10^{-23}$\,cm$^{2}$ & \multicolumn{2}{c}{Dust cross section in wind}  \\
    $\delta$ & $3\times10^{-4}$ & \multicolumn{2}{c}{Dust-to-gas mass ratio in wind}\\
    $f_{\textrm{PAH}}$ & 1.0 & \multicolumn{2}{c}{PAH abundance relative to ISM}\\        
    $\delta_{\textrm{PAH}}$ &$2.6\times10^{-2}$& \multicolumn{2}{c}{PAH-to-dust mass ratio}\\        
   \hline
   \\    
   \hline
   \hline   
   {Other} \\
   \hline
   $\zeta $ & $5\times10^{-17}$\,s$^{-1}$ & \multicolumn{2}{c}{Cosmic ray ionisation rate}\\
   & &  \multicolumn{2}{c}{{e.g. \cite{1999ApJ...512..724B}}}\\
\hline
\end{tabular}
\end{table}

The details of the PDR calculations themselves are given in all of our series of papers on 1D calculations, the latest of which was presenting the \textsc{fried} grid of mass loss rates \citep{2018MNRAS.481..452H}. Nevertheless, the initial abundances are also summarised in Table \ref{PDRGuts} of this paper.  We use the same reduced PDR network of 33 species and 330 reactions. {This has been  tailored to give temperatures accurate to within around 10\,per cent of the UMIST 2012 chemical network database of 215 species and over 3000 reactions \citep{2013A&A...550A..36M}}. The chemistry and thermal balance are solved iteratively until convergence. 
The main coolants are lines from C, C+, O and CO, with some additional contribution from the dust. Heating processes include photoelectric heating from atomic layers of polycyclic aromatic hydrocarbons, C ionisation, H$_2$ formation and photodissociation, FUV pumping, cosmic rays, turbulent and chemical heating and gas-grain collisions. 

The only difference to the microphysics in these calculations compared to those of the \textsc{fried} grid is that we use an interstellar medium (ISM)-like polycyclic aromatic hydrocarbon (PAH)-to-dust mass ratio of $2.6\times10^{-2}$, an order of magnitude higher than that used in the \textsc{fried} grid. PAHs are important, possibly being the dominant heating mechanism in the PDR \citep[see Figure 2 of][which summarises all of the heating and cooling processes of the code]{2016MNRAS.457.3593F} but the PAH abundance in discs is very uncertain \citep{2006A&A...459..545G, 2010ApJ...714..778O, 2011ApJ...727....2P}. In prior work we chose lower values of the PAH abundance to be conservative as to the mass loss rate, but here the higher (ISM like) value leads to a warmer wind that can help achieve a steady state more quickly and hence at less computational cost. Here we are most interested in comparing 1D and 2D models, so the absolute mass loss rate is less important.  We note that any 1D models in this paper are also bespoke models rather than being taken from the \textsc{fried} grid, and hence are also tailored to have the same ISM-like PAH abundance. 

{Finally we note that the microphysics in these calculations is only accurate in the photodissociation regime. At deeper, colder, denser layers within the disc that are extremely optically thick to the UV the chemistry becomes much more complicated and is not captured by our model \citep[see e.g.][]{2010ApJ...722.1607W, 2016A&A...586A.103W, 2017ApJ...843L...3C, 2017A&A...607A..41K, 2019MNRAS.484.1563W}. Although this is unimportant for computing the dynamical evolution of the disc (since such zones are cold regardless of the chemical complexity) we do not attempt to properly capture the composition deep within the disc. Effects such as freeze out upon, for example, the CO abundance distribution can be retrospectively applied for predicting observables.  }

\subsection{Further details}
As discussed above, our calculations involve iteratively solving hydrodynamic and PDR updates via operator splitting. At this stage we are only interested in {the eventual} steady state wind solutions and the associated mass loss rate, so we only do a PDR update every $N$ timesteps. {In practice we start off with $N$ being $\sim10^4$, but as the calculation approaches its final steady state increase $N$ to even larger values ($\sim10^6$).} This means that the pathway to achieving the steady state cannot be studied in detail, but drastically reduces our computational expense. This is the approach we have historically used in our 1D models, for which we confirmed that the frequency with which these updates are made does not affect the final solution \citep{2016MNRAS.463.3616H}. {Note that our models assume chemical and thermal equilibrium}. In section 5.2 of \cite{2016MNRAS.463.3616H} we supported this assumption by showing that the thermal timescale is much shorter than the flow timescale.

\begin{table*}
    \centering
    \begin{tabular}{c c c c c c c c c c}
    \hline
    Model    & $M_*$  & $M_d$ & $R_d$ & $\Sigma_{1\textrm{AU}}$  & UV & $R_{\textrm{crit}}$ (1D) & $R_{\textrm{grid}}$  \\
    ID	& ($M_\odot$) & ($M_{\textrm{jup}}$) & (AU) & (g\,cm$^{-2}$)  & $G_0$ & (AU) & (AU)  \\
    \hline 
    A&1 &10 &100 & 134 & 1000 & 207 & 385  \\
    B&1 &20 &200 & 134 & 1000 & 368 & 770  \\
    C&1 &30 &300 & 134 & 1000 & 380 & 1154  \\       
    \hline
    \end{tabular}
    \caption{A summary of our model parameters. Columns are, from left to right: the model ID, stellar mass, disc mass, disc outer radius, surface density at 1AU, ambient UV field, location of the critical radius in an equivalent 1D model and the grid size. }
    \label{tab:parameters}
\end{table*}

\subsection{Validation}
\textsc{torus-3dpdr} has been benchmarked against 1D semi-analytic solutions in the context of externally irradiated discs in \cite{2016MNRAS.463.3616H}. The code was also validated against a series of dynamical and radiation hydrodynamic tests in \cite{2012MNRAS.420..562H}, \cite{2015MNRAS.448.3156H} and \cite{2015MNRAS.453.1324B}. The PDR components have also been checked against the \cite{2007A&A...467..187R} benchmarks in \cite{2015MNRAS.454.2828B}

The \textsc{healpix} scheme used by \textsc{torus-3dpdr} is known to work in 3D applications, however it has been modified for this work to utilize the cylindrical symmetry of our simulation grid when propagating through 3D space. To check this modified scheme we computed the UV field impinging upon a sphere embedded in a low density medium using both the 3D and 2D cylindrical scheme, ensuring that the same UV field in the sphere resulted.

\subsection{Disc construction/boundary conditions}
\label{sec:disc}
In these calculations we define a disc that is imposed and hence acts as a boundary condition and isn't allowed to dynamically evolve (in the first instance we are interested in obtaining steady state solutions to compare with 1D models). For this purpose  we consider a truncated power law for the imposed disc surface density profile
\begin{equation} 
	\Sigma(R) = \Sigma_{1\textrm{AU}}\left(\frac{R}{\textrm{AU}}\right)^{-1}
\end{equation}
where for a disc of a given size $R_d$ and mass $M_d$ the surface density normalization is
\begin{equation}
	\Sigma_{1\textrm{AU}} = \frac{M_d}{2\pi R_d \textrm{AU}}.
\end{equation}
We impose the disc up to one scale height
\begin{equation}
	H=\frac{c_s}{\Omega}
\end{equation}
where $c_s$ is the sound speed and $\Omega$ the Keplerian angular velocity. This scale height is computed assuming an isothermal disc temperature of 20\,K. We \textit{do} allow the disc itself to be subsequently heated in the PDR calculation, but do not  then adjust the height above the mid-plane to which the disc is imposed (i.e. we always impose it up to 1 scale height as though the disc temperature were 20K). We retrospectively checked the disc mid-plane temperature profile and it is in the range 20--23\,K in the bulk of the outer disc, excluding the outermost few cells which can be heated to higher temperatures. Where the external radiation field becomes extremely weak the source star may set the temperature through passive irradiation, which we assume is of the form
\begin{equation}
	T_* = 100\left(\frac{R}{\textrm{AU}}\right)^{-1/2}. 
\end{equation}
We take the maximum of the external radiation and stellar-induced temperatures, with a floor value of 20\,K. Note that the assumed temperature of 20\,K in setting the imposed scale height is accurate to within a few degrees in the mid-plane except for the outermost $\sim8$ cells, though of course at higher $z$ the disc gets significantly warmer, as we will illustrate in section \ref{ref:denstemp}.

\subsection{Model parameters}
We consider 3 calculations in this paper, the parameters of which are summarised in table \ref{tab:parameters}. At this stage we are considering only a 1\,M$_\odot$ star, and all of our models consider a $10^3$\,G$_0$ field. We do however vary the disc mass and radius such that the surface density normalisation is constant. Of course there will likely be interesting variations in the nature of external disc photoevaporation in different stellar mass and UV regimes that will need to be explored in future work. 

In 1D models the required size of the grid is tightly constrained. A unique solution for a steady  transonic flow satisfies a criticality condition \citep[simultaneous
vanishing of terms in the combined momentum and continuity equations, e.g.][]{1965SSRv....4..666P, 2016MNRAS.460.3044C} at some point in the flow, given by that at which
\begin{equation}
	\frac{v^2\mu m_{\textrm{H}}}{k_{\textrm{B}}} - T - n \frac{\textrm{d}T}{\textrm{d}n} \geq 0.
	\label{equn:rCrit}
\end{equation}
is first satisfied for velocity $v$, mean particle mass $\mu$, temperature $T$ and number density $n$ \citep{2016MNRAS.457.3593F}. 1D models will only
converge on this solution if the critical point is contained within the grid. Our 2D grid is chosen to be larger than this critical radius for the 1D models in each case.

The 3 main calculations are predominantly presented using a fixed $256\times256$ cell cylindrical $R-z$ grid and the lowest level of \textsc{healpix} refinement (12 rays per cell, {propagated in 3D utilising the axisymmetry of our model}). {In section \ref{sec:streamlineconvsec} we also explore convergence, running  $128\times128$ and $64\times64$ cell versions, and for the lowest resolution case also using the next 2 levels of \textsc{healpix} refinement (48 and 192 rays per cell), though we are unable to practically increase both the spatial and \textsc{healpix} resolution further at this stage. }

\subsection{Computing the mass loss rate}
\label{sec:mdotDescription}

One of the primary quantities of interest from these calculations is the mass loss rate, which is important for the dynamical evolution and lifetime of the disc.  In 1D calculations the mass loss rate is computed following \cite{2004ApJ...611..360A} using
\begin{equation}
	\dot{M} = 4\pi R^2 \rho \dot{R} \mathcal{F},
	\label{equn:mdot1Da}
\end{equation}
where $\mathcal{F}$ is the fraction of solid angle subtended by the disc outer edge
\begin{equation}
	\mathcal{F} = \frac{H_d}{\sqrt{H_d^2+R_{d}^2}}. 
	\label{equn:mdot1Db}	
\end{equation}

In our new 2D calculations the mass loss rate is computed through a spherical surface of radius $r$ at the end of the calculation in a postprocessing step as follows. We generate a number of evaluation
points on a circular contour in the ($R,z$) plane of radius $r$ centred on the origin. From any given evaluation point we draw a tangent to this circle and
calculate the length of this tangent that lies within the cell
that contains it. We ensure that each $i^{th}$ cell only contributes once to the mass loss rate estimate, which is given by
\begin{center}
	\begin{equation}
		\dot{M_i} = \left\{
    		\begin{array}{l l}
    			2\times2\pi{R}\Delta{l}\rho_i\sqrt{v_{R,i}^2+v_{z,i}^2}, & v_R > 0 \,\,\,\, \textrm{or} \,\,\,\, v_z >0 \\
			 & \\
    			0 & \textrm{otherwise} \\
    		\end{array} \right.
	\label{equn:discreteMdot}
	\end{equation}
\end{center}
where $v_{R,i}$ and $v_{z,i}$ are the $R$ and $z$ components of the velocity and the preceding factor of 2 arises because the models in this paper are reflective about the mid-plane. 

\begin{figure}
	\hspace{-0.3cm}
	\includegraphics[width=9.5cm]{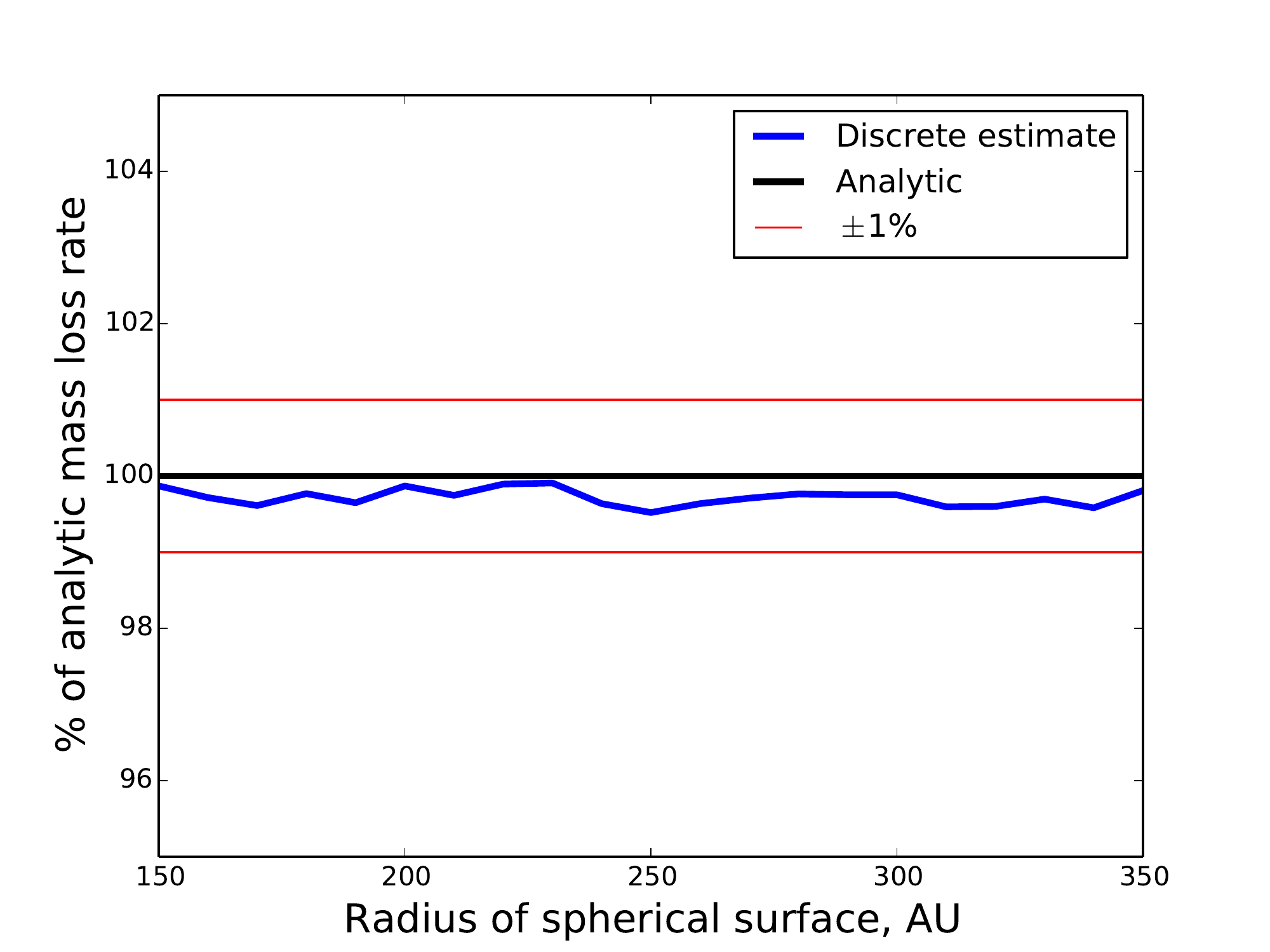}
	\caption{The mass loss rate calculated on our discretised grid for a spherically diverging flow, compared with the analytic mass loss rate. In this example our approach is accurate to with 1\,per cent (the red horizontal lines) and is a conservative measure, only ever underestimating the mass loss rate. }
	\label{fig:mdotTesting}
\end{figure}

We tested this scheme for a spherically symmetric diverging flow of constant velocity and inverse square density profile (hence a constant ``mass loss rate'' as a function of radius). For a spherically diverging velocity of $10^{-3}$\,km\,s$^{-1}$ and density profile of the form
\begin{equation}
	\rho = 10^{-20}\left(\frac{R}{100\,\textrm{AU}}\right)^{-2}\,\textrm{g}\,\textrm{cm}^{-3}
\end{equation}
the analytic mass loss rate is $4.46\times10^{-13}$\,M$_\odot$\,yr$^{-1}$. The mass loss rate we obtain by summing equation \ref{equn:discreteMdot} over a spherical surface, as a function of the radius of that spherical surface, is given in Figure \ref{fig:mdotTesting}. This was for the case of a fixed 385\,AU 2D cylindrical grid with $256^2$ cells. Our discretised estimates agree with the analytic value to within 1\,per cent for all spherical radii through which we compute the mass loss rate. 

\begin{figure*}
	\hspace{-1.cm}
	\includegraphics[width=2.2cm]{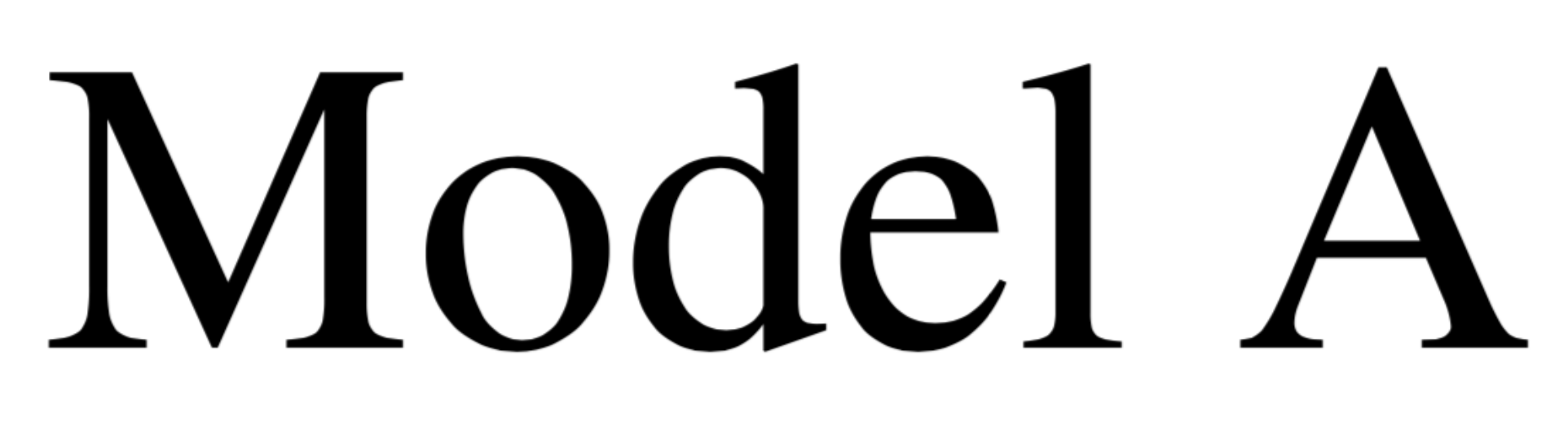}
	\hspace{4cm}	
	\includegraphics[width=2.2cm]{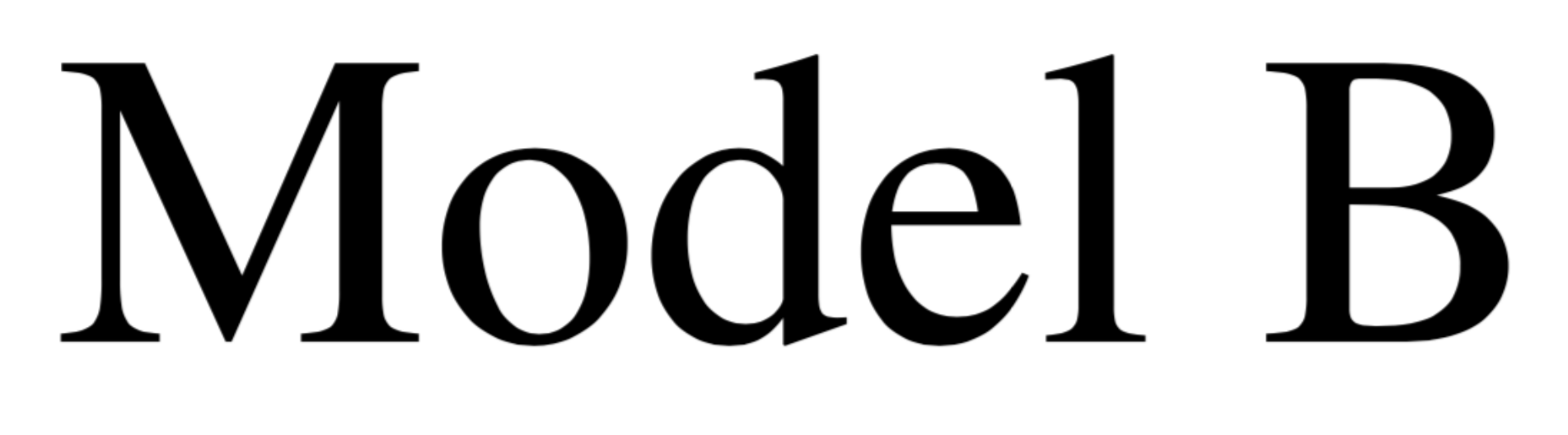}
	\hspace{4cm}	
	\includegraphics[width=2.2cm]{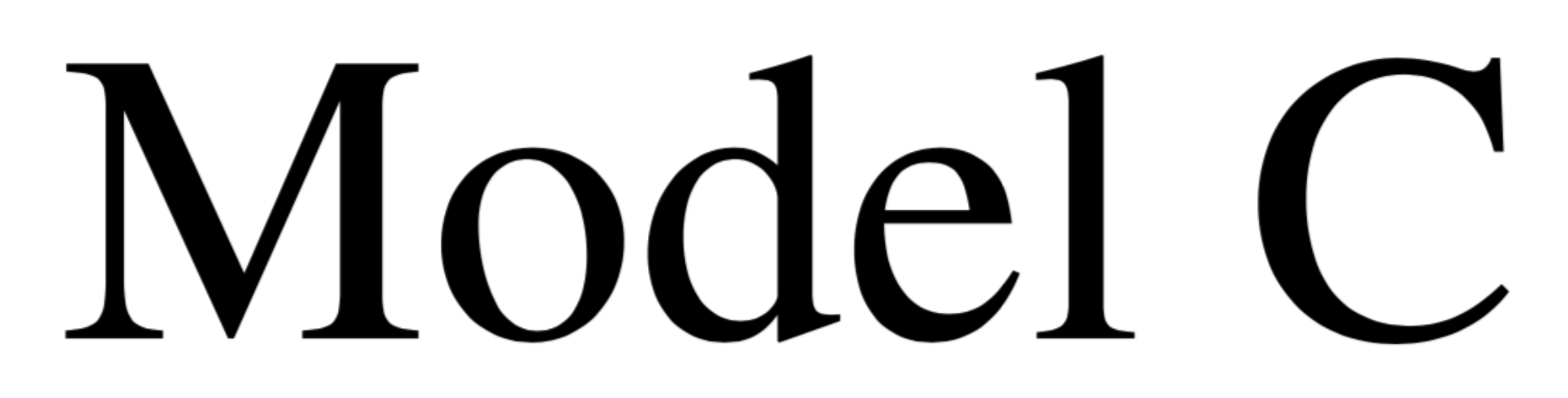}	

	\hspace{-0.8cm}
	\includegraphics[width=6.cm]{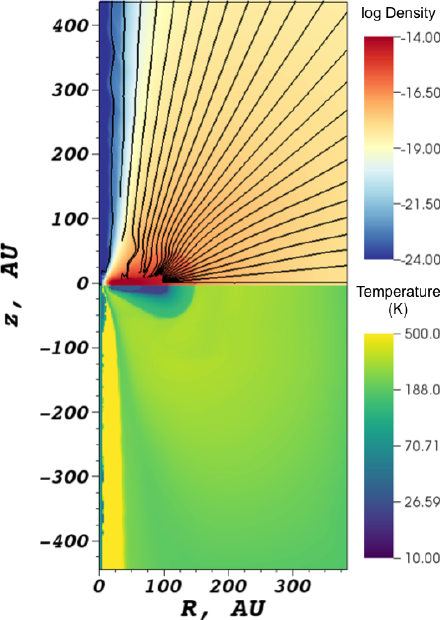}
	\includegraphics[width=6.cm]{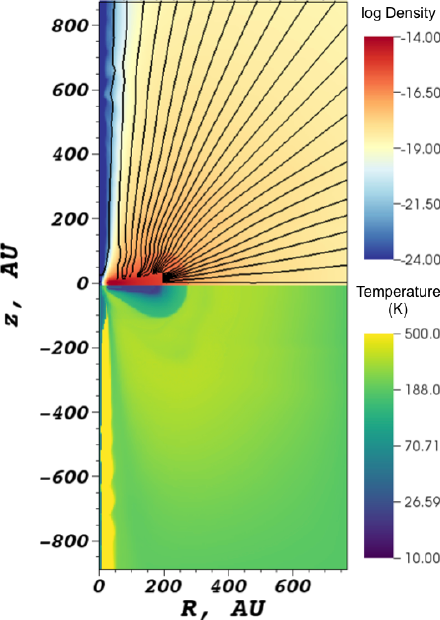}
	\includegraphics[width=6.cm]{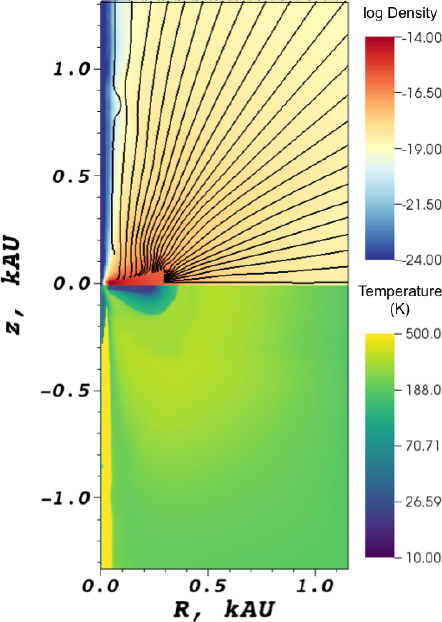}	
	\hspace{0.cm}			
	\caption{A summary of the density structure, streamlines and temperature distribution in our models. The upper frame of each panel shows the $\log_{10}$ density, with velocity streamlines overlaid. The lower frame of each panel is the temperature distribution, reflected about the disc mid-plane.}
	\label{fig:rhoVT}
\end{figure*}

\section{Dynamics and chemistry of multidimensional external disc photoevaporation}
\label{ref:denstemp}
The analysis of our models proceeds as follows. In section \ref{sec:overview} we make an initial overview of the steady state flow, thermal and kinematic structure. In section  \ref{sec:flowOrigin} we then quantify the location in the disc from which the mass loss originates. With an understanding of the flow structure, we then compute and compare 1D and 2D mass loss rates in \ref{sec:mdots}. We then discuss convergence of our models in \ref{sec:convergence} and finish with a first look at the chemical composition of externally irradiated discs in 2D in \ref{sec:chemistry}.

\begin{figure*}
	\hspace{-1.cm}
	\includegraphics[width=2.2cm]{modelA.pdf}
	\hspace{4cm}	
	\includegraphics[width=2.2cm]{modelB.pdf}
	\hspace{4cm}	
	\includegraphics[width=2.2cm]{modelC.pdf}

	\hspace{-1.cm}	
	\includegraphics[width=6.15cm]{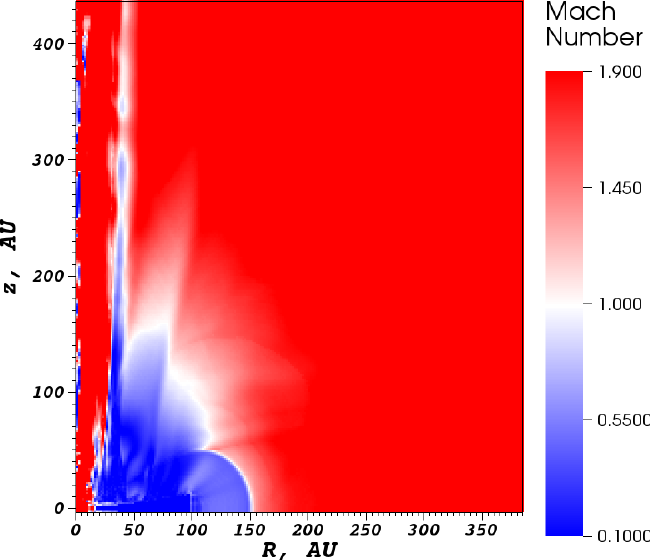}
	\hspace{-0.1cm}	
	\includegraphics[width=6.15cm]{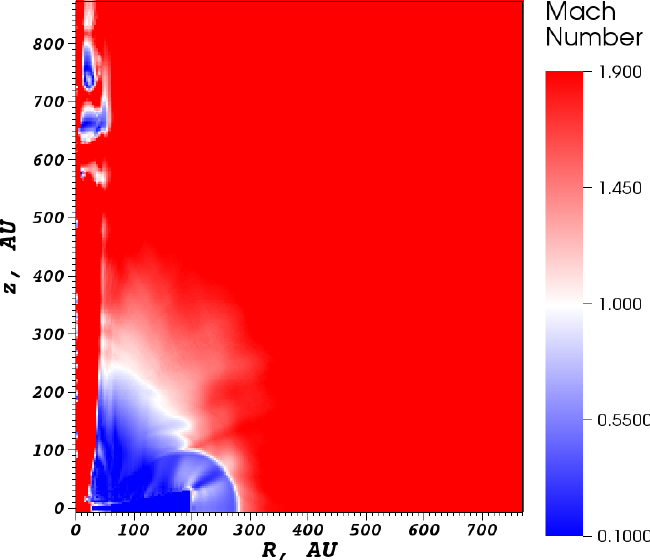}	
	\hspace{-0.1cm}		
	\includegraphics[width=6.15cm]{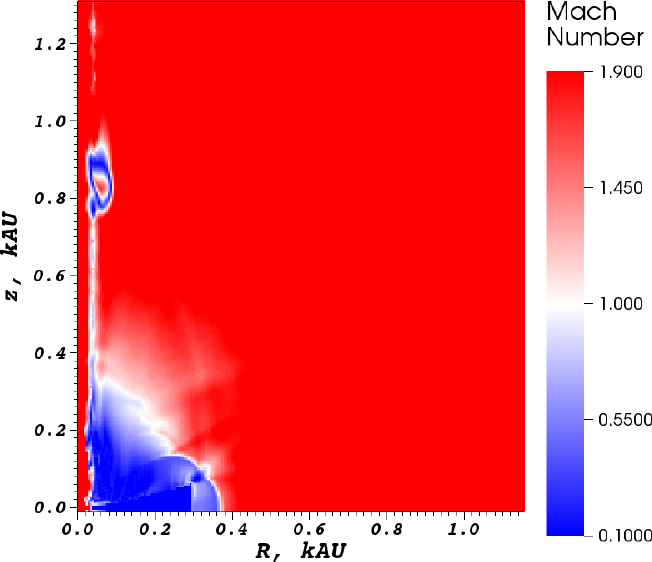}

	\caption{Maps of the Mach number in our models. The colourbar is chosen such that subsonic components of the flow are blue and supersonic components are red. The sonic surface is approximately denoted by the white parts of the image. }
	\label{fig:mach}
\end{figure*}

\subsection{Initial overview of flow structure}
\label{sec:overview}
We begin with an initial broad overview of the steady state flow structure in our models. Figure \ref{fig:rhoVT} shows the density and temperature structure, as well as a selection of flow streamlines. Figure \ref{fig:mach} also shows the Mach number, colour coded such that subsonic flow is blue, supersonic red and the approximate sonic surface is white. In all cases  the cold outer disc is surrounded by a warm ($\sim50-150$\,K) layer of subsonic material of typical density $\sim10^8$\,cm$^{-3}$.  The edge of this ``halo''\footnote{We now refer to the disc region that we impose as ``the disc'' and any CO bearing region outside of this as ``the halo'', motivated by the terminology used in the interpretation of IM Lup \citep{2016ApJ...832..110C, 2017MNRAS.468L.108H}} corresponds to the attainment of an optical depth $\sim1$, which coincides with the surface of photodissociation of CO (see section \ref{sec:chemistry}) and the transition to a hotter ($\sim200-350$\,K) supersonic flow. Above the disc surface, interior to about half of the disc radius (a region that contributes negligibly to the mass loss: see section  \ref{sec:flowOrigin}) the flow is subsonic up to a large height. At small $R$ a hot ($\sim1500$\,K in the $10^3$\,G$_0$ cases) low density ($10^{-3}$\,cm$^{-3}$) cavity is blown along the $z-$axis which is subject to small scale changes over time due to small scale shearing instabilities. The rest of the flow and hence mass loss rate is steady, which we will illustrate further in section \ref{sec:convergence}. 

The streamline morphology, as illustrated in the upper panels of Figure \ref{fig:rhoVT} as well as in more detail for model B in Figure \ref{fig:Bstreams}, is similar in each of our three cases. Streamlines emanating from the disc surface leave vertically, before turning over and becoming quasispherical at larger radii. Streamlines emanating from the disc outer edge flare such that the expansion is greater than spherical out to approximately 1.5 disc radii, before again turning over to become more spherical.

\subsection{Where is mass lost from the disc?}
\label{sec:flowOrigin}
Before discussing the total mass loss rates in our models we first discuss the radial location in the disc from which the bulk of the mass loss originates, as well as how it propagates in the wind. This is a prerequisite to understanding any differences in the mass loss rate between the 1D and 2D models.  

The gravitational radius $R_g$ is that at which the sound speed is equal to the escape velocity (or equivalently the thermal energy equals the binding energy) and hence matter is unbound 
\begin{equation}
	R_g = \frac{GM_*\mu m_H}{k_B T}. 
\end{equation}
\citep[e.g.][]{1994ApJ...428..654H, 1998ApJ...499..758J}. Scenarios where $R_g$ is greater than the disc radius are referred to as sub-critical, and a key assumption of the 1D models of externally evaporating discs in the sub-critical regime is that mass loss is dominated by that from the disc outer edge. 1D models then  compute the mid-plane flow structure and calculate a total mass loss rate assuming a spherically diverging wind (as in equations \ref{equn:mdot1Da} and \ref{equn:mdot1Db}). 

\begin{figure}
	\hspace{-0.4cm}
	\includegraphics[width=9.5cm]{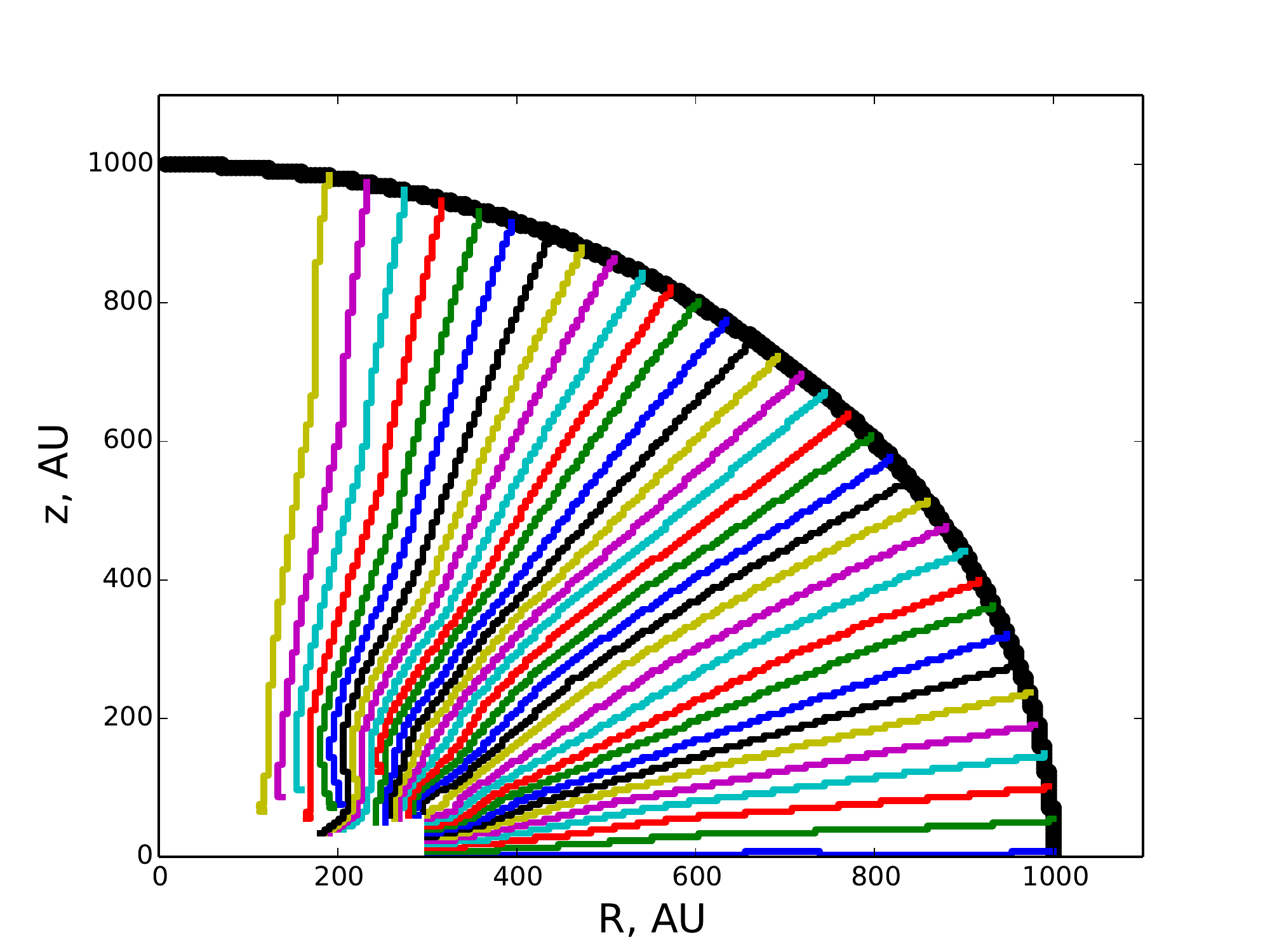}
	\caption{The cross section of the spherical surface through which the mass loss rate is calculated (black points) for the case of model C. Included are a selection of streamlines from this surface, traced back towards the disc.}
	\label{fig:Bstreams}
\end{figure}

In our models, which are all sub-critical (we verified this by checking maps of $R_g/R_d$) we can compute exactly where the mass loss originates from in the disc and compare with the 1D assumptions above. Recall from the description of our method in \ref{sec:mdotDescription} that we compute the mass loss rate in the 2D models though a spherical surface. From each grid cell on this surface (where we know the local contribution to the total mass loss rate) we trace a streamline back to its origin at the disc. An illustration of this for an example spherical surface and handful of streamlines in model C is given in Figure \ref{fig:Bstreams}. 

\begin{figure}
	\hspace{-0.1cm}
	\includegraphics[width=9.7cm]{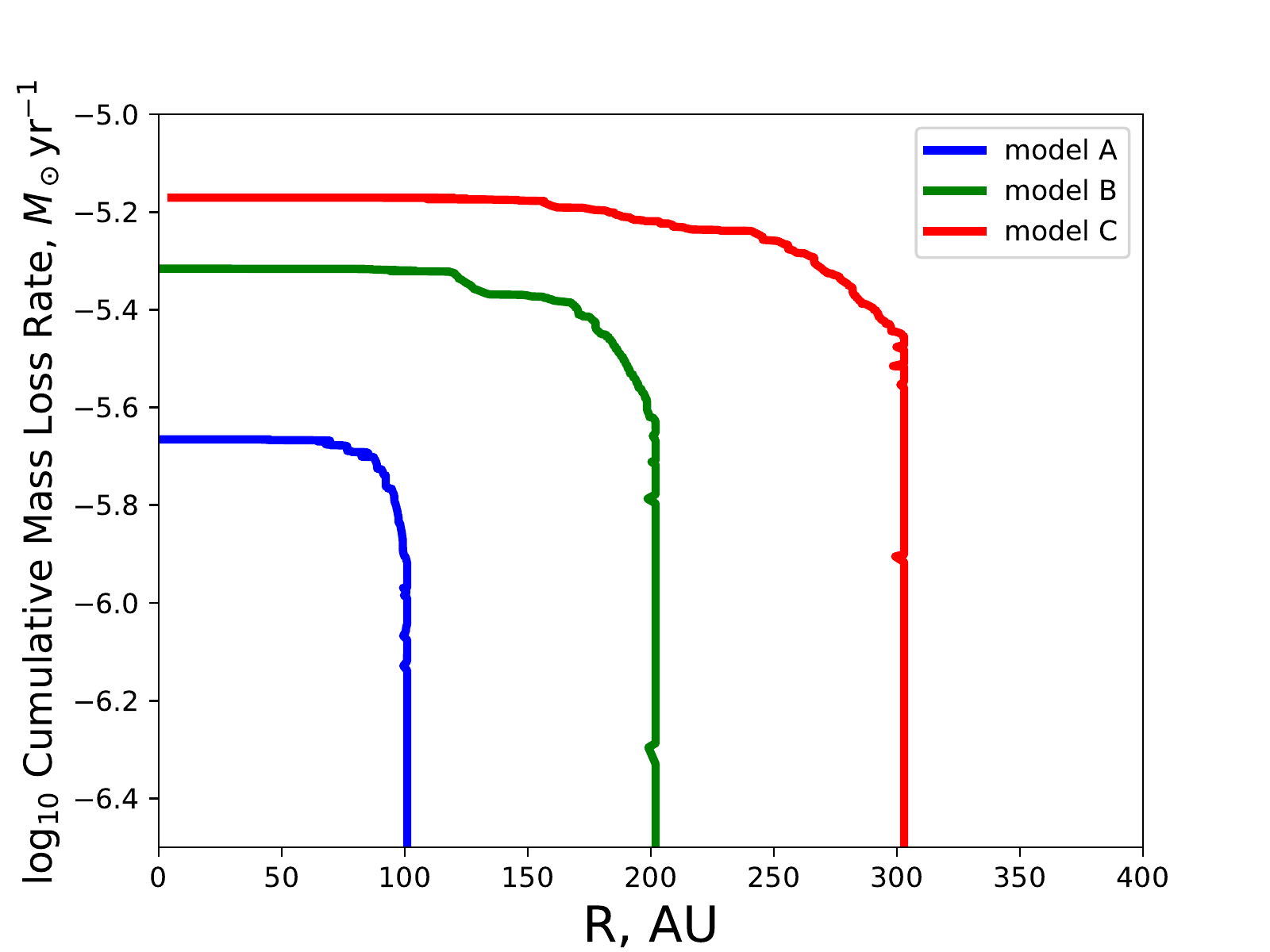}

	\vspace{0.1cm}
	\hspace{-0.1cm}	
	\includegraphics[width=9.7cm]{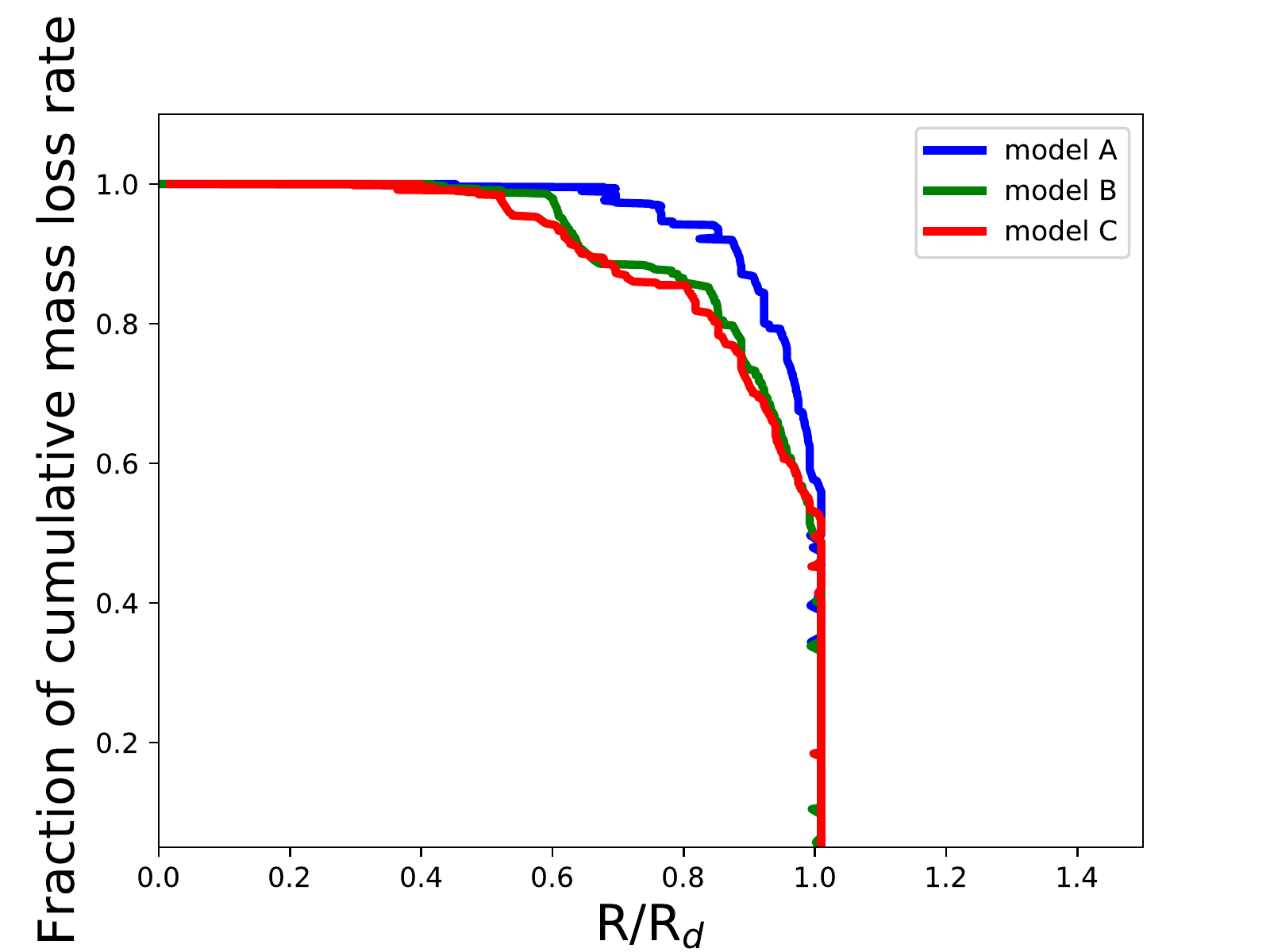}
	\caption{The cumulative mass loss rate as a function of the radial origin of streamlines. The upper panel is the absolute mass loss rate and distance and the lower panel is normalised to the total mass loss rate and disc outer radius. Note that the disc outer radii are 100, 200 and 300\,AU in models A, B and C respectively.  }
	\label{fig:streamOrigins}
\end{figure}

Using this streamline tracing approach, in the upper panel of Figure \ref{fig:streamOrigins} we plot the cumulative mass loss rate through the spherical surface as a function of the $R-$coordinate at which the streamline originates from the disc. The lower panel shows the same information, normalised to the total mass loss rate and disc outer radius. These plots demonstrate that \textit{in all cases} around 50\,per cent of the mass loss rate comes directly from the disc outer edge. Although some mass loss does come from the surface, \textit{the mass loss rate is entirely set in the outer half of the disc and is mostly from within about 10\,per cent of the disc outer edge}. That is, in the lower panel of Figure \ref{fig:streamOrigins} the fraction of cumulative mass loss is 1 by $R/R_d=0.5$. Unsurprisingly larger (less bound) discs lose mass from a larger fraction of the disc surface. Overall the assumption from 1D models that the mass loss is predominantly from the disc outer edge therefore seems reasonable to within a factor two or so. We will compare the actual 1D and 2D mass loss rates in section \ref{sec:mdots}.

\begin{figure*}
	\hspace{-1.cm}
	\includegraphics[width=2.2cm]{modelA.pdf}
	\hspace{4cm}	
	\includegraphics[width=2.2cm]{modelB.pdf}
	\hspace{4cm}	
	\includegraphics[width=2.2cm]{modelC.pdf}

	\hspace{-0.8cm}	
	\includegraphics[width=6.42cm]{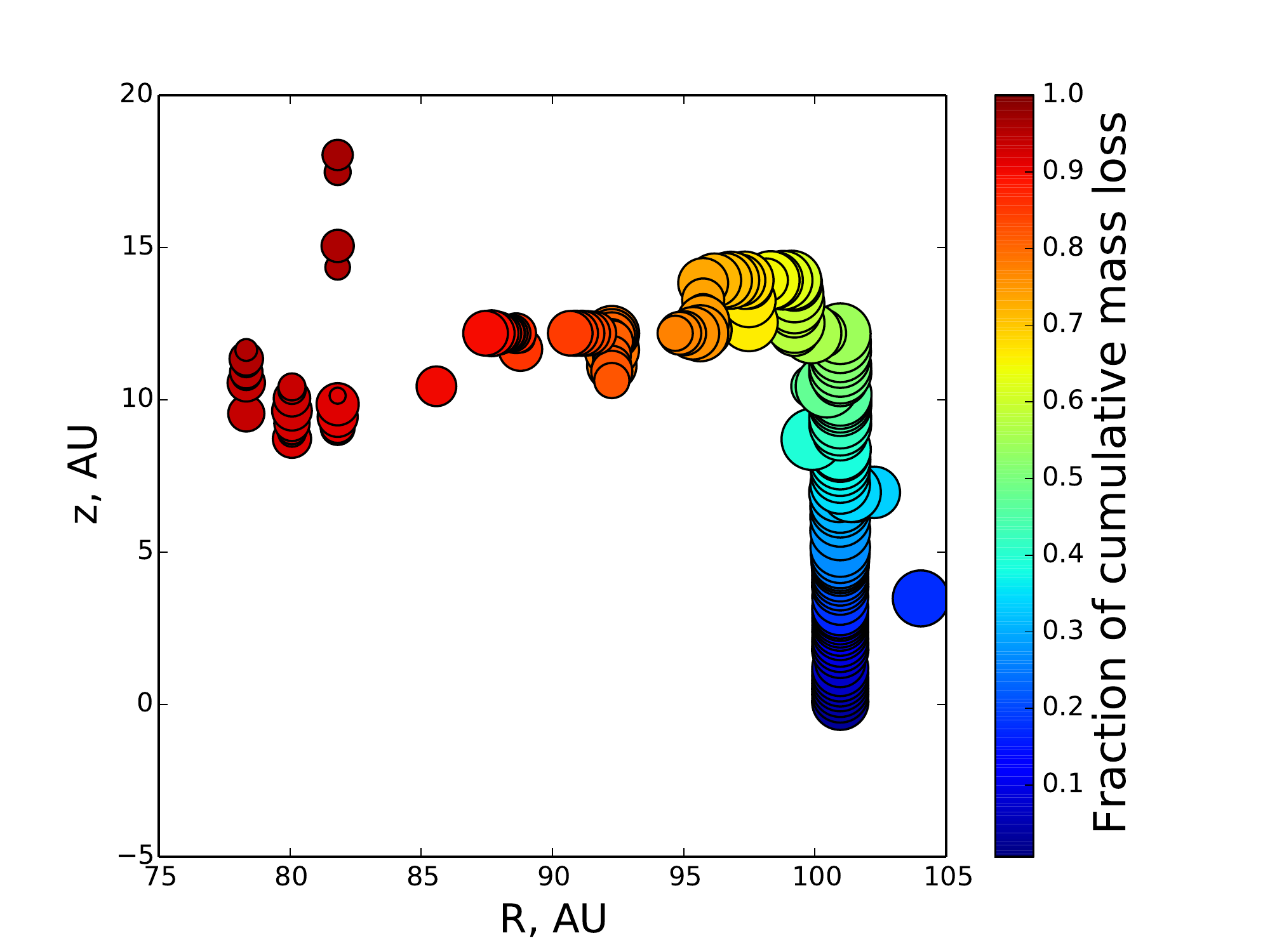}
	\hspace{-0.7cm}		
	\includegraphics[width=6.42cm]{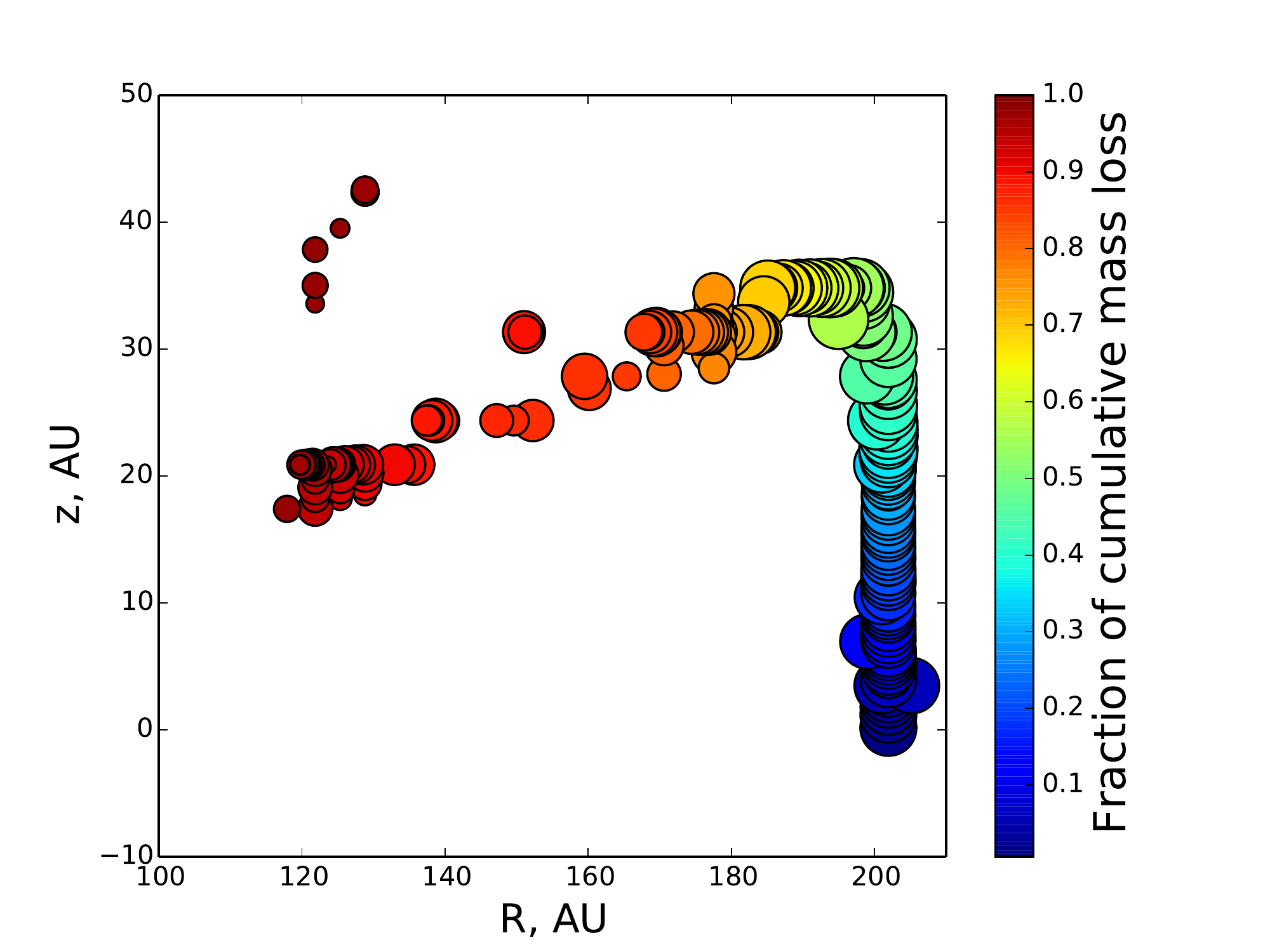}	
	\hspace{-0.7cm}		
	\includegraphics[width=6.42cm]{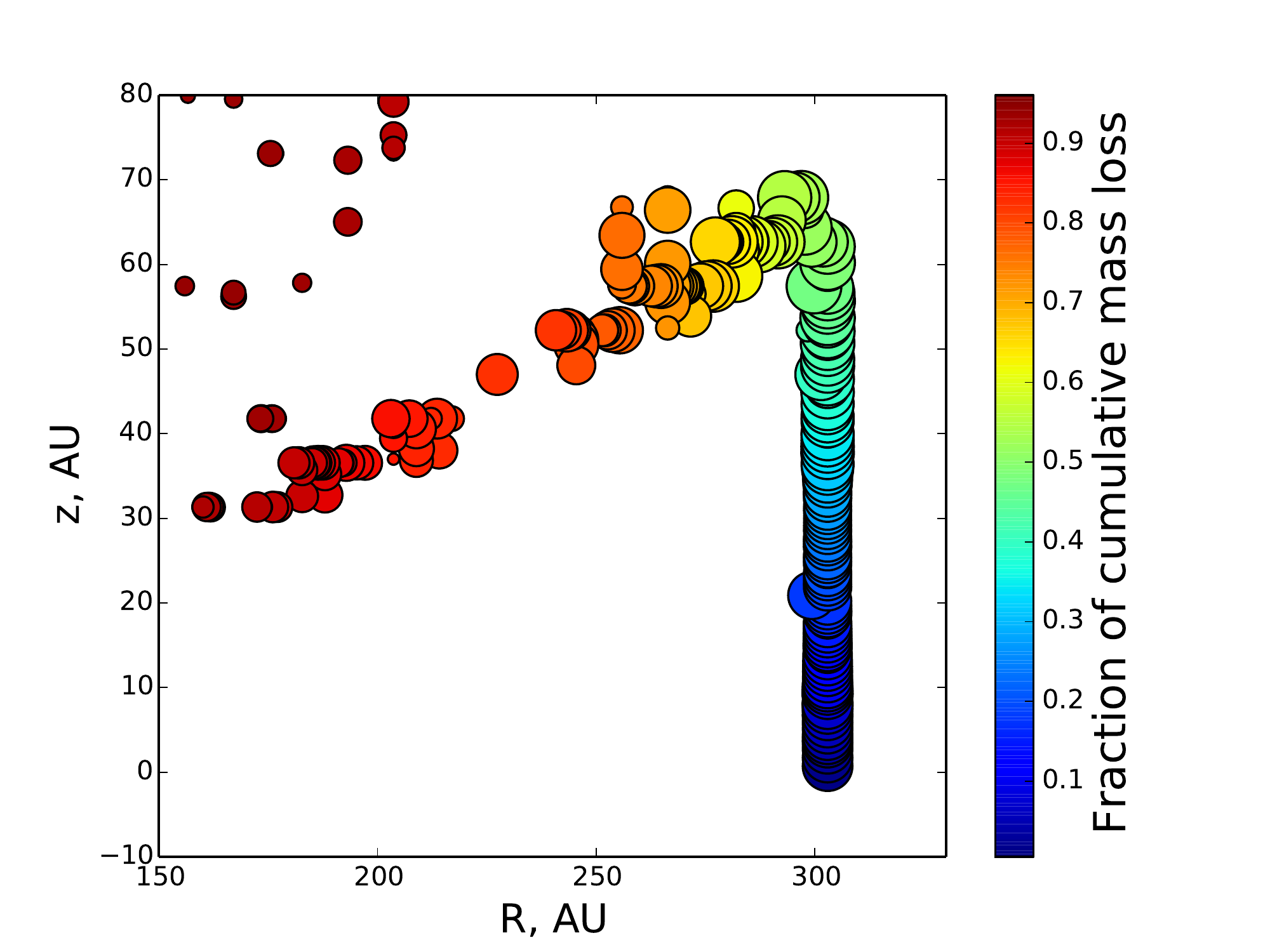}

	\caption{The mass loss rates in our models are computed through a spherical surface. The points in these plots are where streamlines from this surface trace back to on the disc itself. The points are colour coded by the cumulative mass loss rate. E.g. 50\,per cent of all mass loss takes place between the dark blue and green points, $\sim$70 percent between dark blue and orange and 100\,per cent between dark blue and dark red. These plots further illustrate that around 50\,per cent of the mass loss is from the disc outer edge, with the rest originating from the disc surface close to the disc outer edge. Points are also scaled in size by the ratio of the mass loss for any given streamline to the total mass loss rate.  }
	\label{fig:streamOriginsScatter}
\end{figure*}

We further illustrate the origin of mass loss in our models in Figure \ref{fig:streamOriginsScatter}. This shows the spatial ($R-z$) location of the streamline end points near the disc outer edge. These points are colour-coded by the cumulative mass loss rate. Starting from the dark blue points, one can see what fraction of the mass loss originates from certain locations. For example, at the top of the disc outer edge the points are green, corresponding to roughly half of the total mass loss rate.

\subsection{Comparing 1D and 2D mass loss rates}
\label{sec:mdots}
The key quantity of interest from external photoevaporation for disc evolutionary models is the mass loss rate. Mass loss rates from 1D models are being used in viscous evolutionary calculations to study disc evolution in different environments, where the role of environment is being predicted to be important \citep[e.g.][]{2007MNRAS.376.1350C, 2013ApJ...774....9A, 2018MNRAS.475.5460H, 2018MNRAS.481..452H, 2017MNRAS.468.1631R, 2018MNRAS.478.2700W, 2018ApJ...867...41S}. It is crucial to confirm the validity (or otherwise) of these mass loss rates in multidimensional models. 

\begin{figure}
	\hspace{-0.4cm}	
	\includegraphics[width=9.5cm]{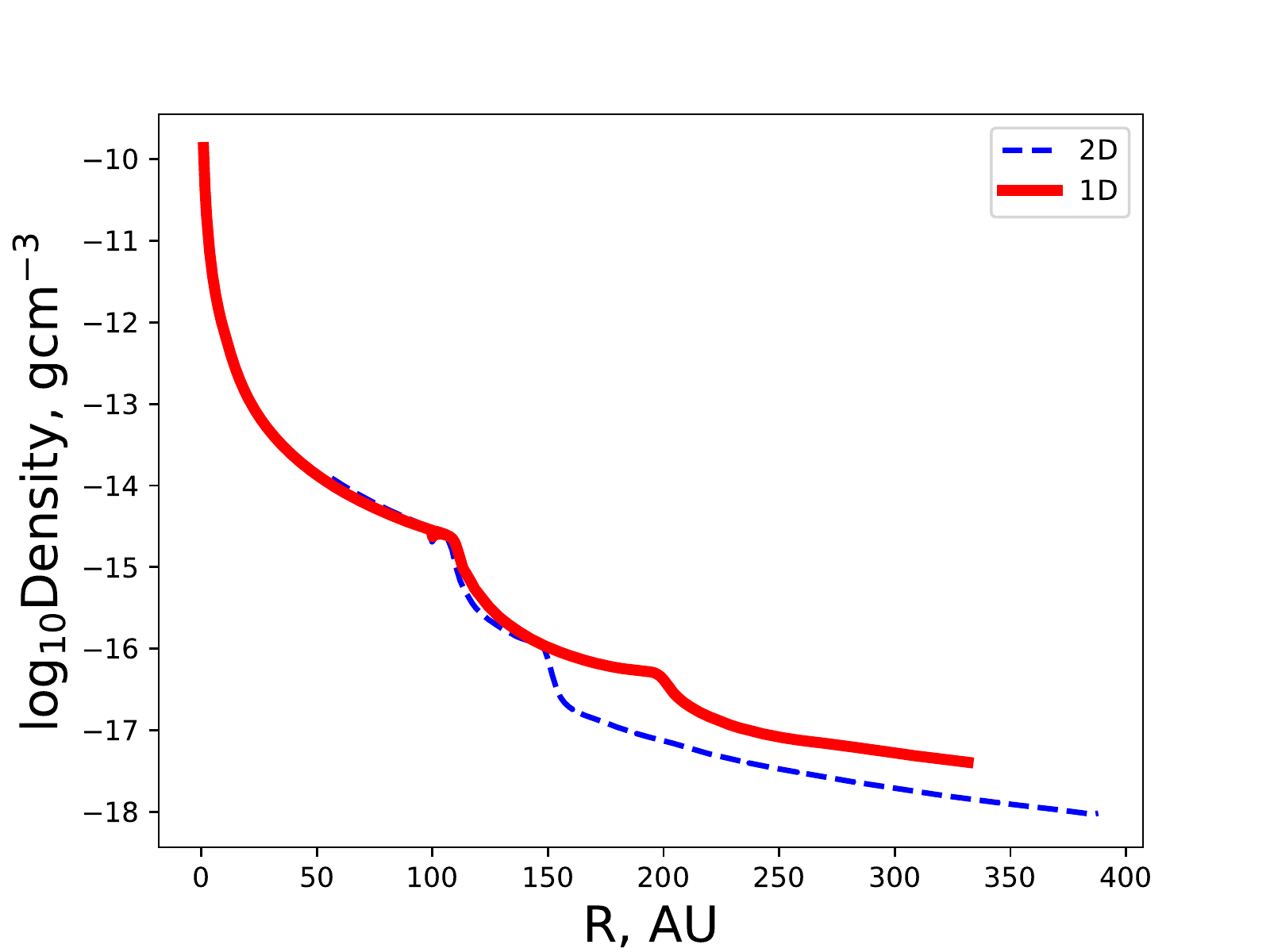}

	\hspace{-0.4cm}	
	\includegraphics[width=9.5cm]{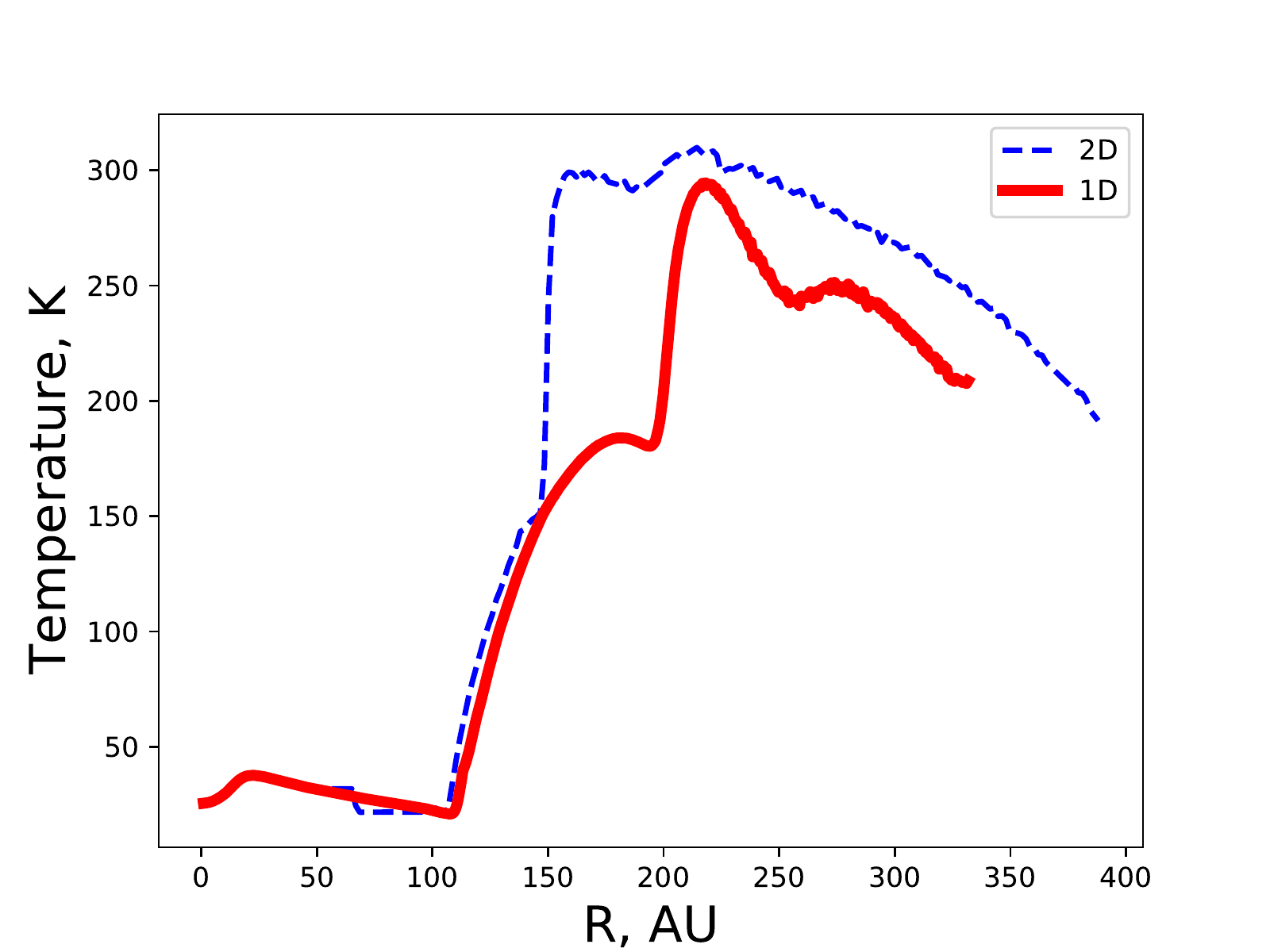}	

	\hspace{-0.4cm}		
	\includegraphics[width=9.5cm]{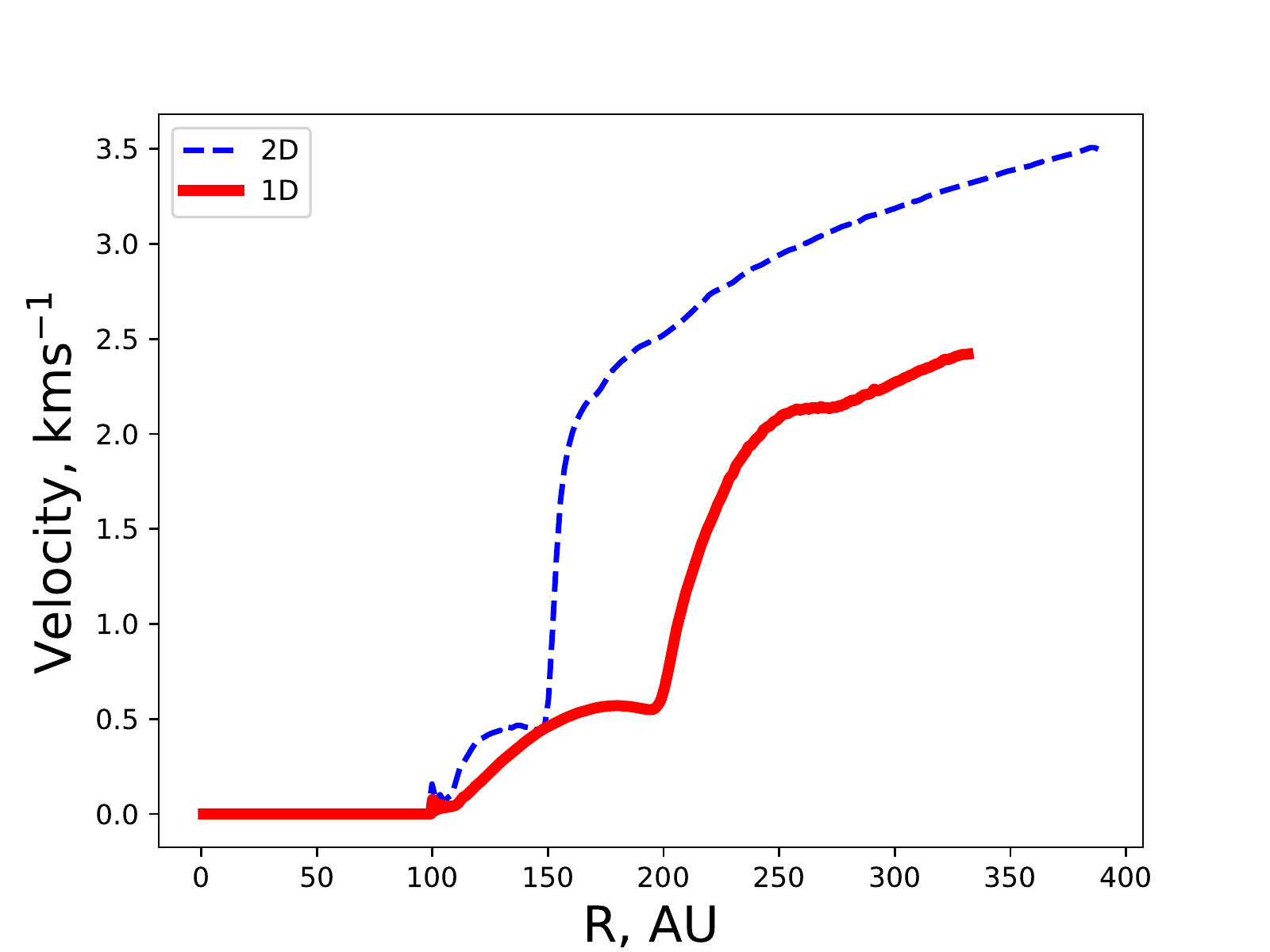}

	\caption{A comparison of  the mid-plane flow variables in the 1D (red) and 2D (blue) versions of model A. The panels are the density, temperature and velocity structure from top to bottom.  }
	\label{fig:modelAMidPlaneFlowVariables}
\end{figure}

We ran 1D counterparts to our 2D models to compare the mass loss rates \citep[e.g. similar to the \textsc{fried} grid models][only with the required higher PAH abundance used here]{2018MNRAS.481..452H}. These 1D models would be defined as having the same disc mass and radius, as well as stellar and irradiating UV field properties as the 2D models. The 1D models do differ in that the scale height at the disc outer edge is allowed to increase due to heating at the disc outer edge (as adopted in the \textsc{fried} grid). Note that in the 2D models we calculate the mass loss rate through a number of spherical surfaces over the supersonic flow and take the average (it should be constant in a perfect steady state and in section \ref{sec:spheresizesec} we will show that this is the case to within $\pm3$\,per cent). For reference, the mid-plane flow variables for model A, which are typical of each of the models, are shown in Figure \ref{fig:modelAMidPlaneFlowVariables}. The 2D models result in more rarefied, faster flows in the supersonic wind along the mid-plane. 

\begin{table}
    \centering
    \begin{tabular}{c c c c c c c c c }
    \hline
    Model ID      & $\dot{M}$  2D  &  $\dot{M}$ 1D &  2D / \\
   	&  ($10^{-6}M_{\odot}$\,yr$^{-1}$) & ($10^{-6}M_{\odot}$\,yr$^{-1}$)  & 1D    \\
    \hline 
    A & {2.2} & 0.6 & 3.7 \\
    B & {4.8}& 3.1  & 1.55 \\
    C &  {6.8} & 5.8 & 1.17 \\
    \hline
    \end{tabular}
    \caption{Mass loss rates 2D and equivalent 1D models, as well as their ratio.  }
    \label{tab:2Dvs1D}
\end{table}

Table \ref{tab:2Dvs1D} compares the total mass loss rate in our 2D models with their 1D analogues. We are probing only a very limited parameter space in this paper so broad generalisations must be treated with caution. However, in all of the models here the mass loss rate is larger in the 2D cases. So the contribution of mass loss from the disc surface and faster flow more than compensates for the 1D models being based entirely on the peak density part of the flow along the mid-plane.

Recall that 1D models have been predicting that external photoevaporation of discs is important for disc evolution (and potentially planet formation) in a wide range of regimes. Our 2D models are suggesting that in fact the 1D mass loss rates are probably underestimates (albeit by factors of order unity) and so the impact is, if anything, likely stronger than previously anticipated.

\subsection{Convergence}
\label{sec:convergence}
We now briefly assess the convergence properties of our models. 

\subsubsection{Sensitivity of mass loss rate to size of spherical surface}
\label{sec:spheresizesec}
\begin{figure}
	\hspace{-0.4cm}
	\includegraphics[width=9.cm]{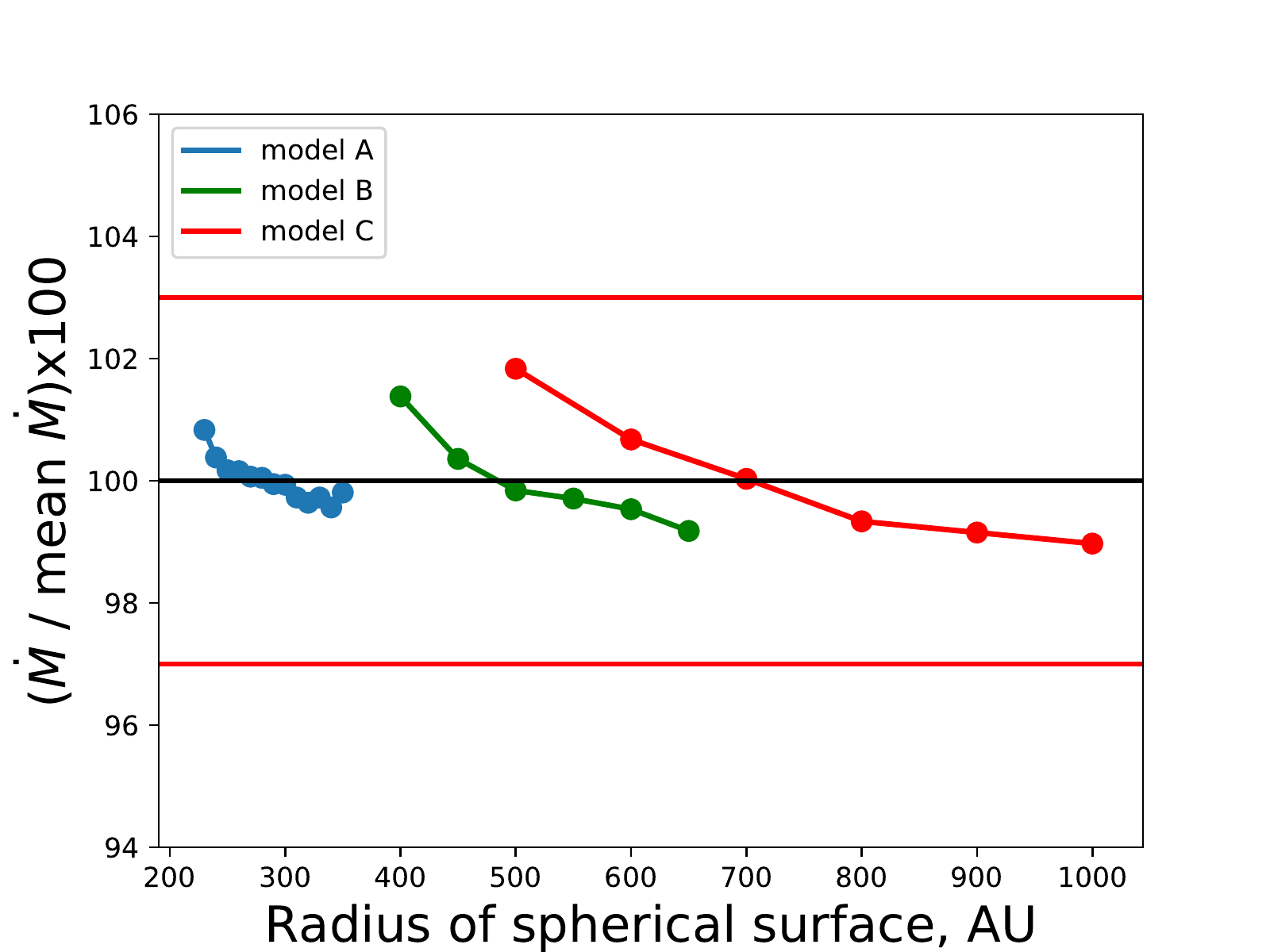}
	\caption{The mass loss measured as a function of spherical surface radius, expressed as the ratio of local to mean mass loss rate.  The red horizontal lines are the 3\,per cent level  }
	\label{fig:varySphere}
\end{figure}

We calculated the mass loss rate in our models through spherical surfaces of different sizes and used the mean. In \ref{sec:mdotDescription} we showed that the variation for a perfect spherically diverging flow is only about 1\,per cent over 200\,AU variation in surface radius. Figure \ref{fig:varySphere} shows the percentage variation of the mass loss rate, relative to the mean as a function of the radius of the spherical surface. Generally the mass loss rate slightly decreases with surface size, but is flattening out at larger radii (as the number of cells contributing to the estimate increases). Overall the variation is never more than  $\pm3$\,per cent. That this is higher than the 1\,per cent level in the perfect case discussed in \ref{sec:mdotDescription} will in part be due to small deviations from a perfectly steady state, which make a bigger perturbation to the total mass loss rate at smaller radii (where there are a smaller number of cells on the surface) than at large radii.

\begin{figure}
	\includegraphics[width=9.2cm]{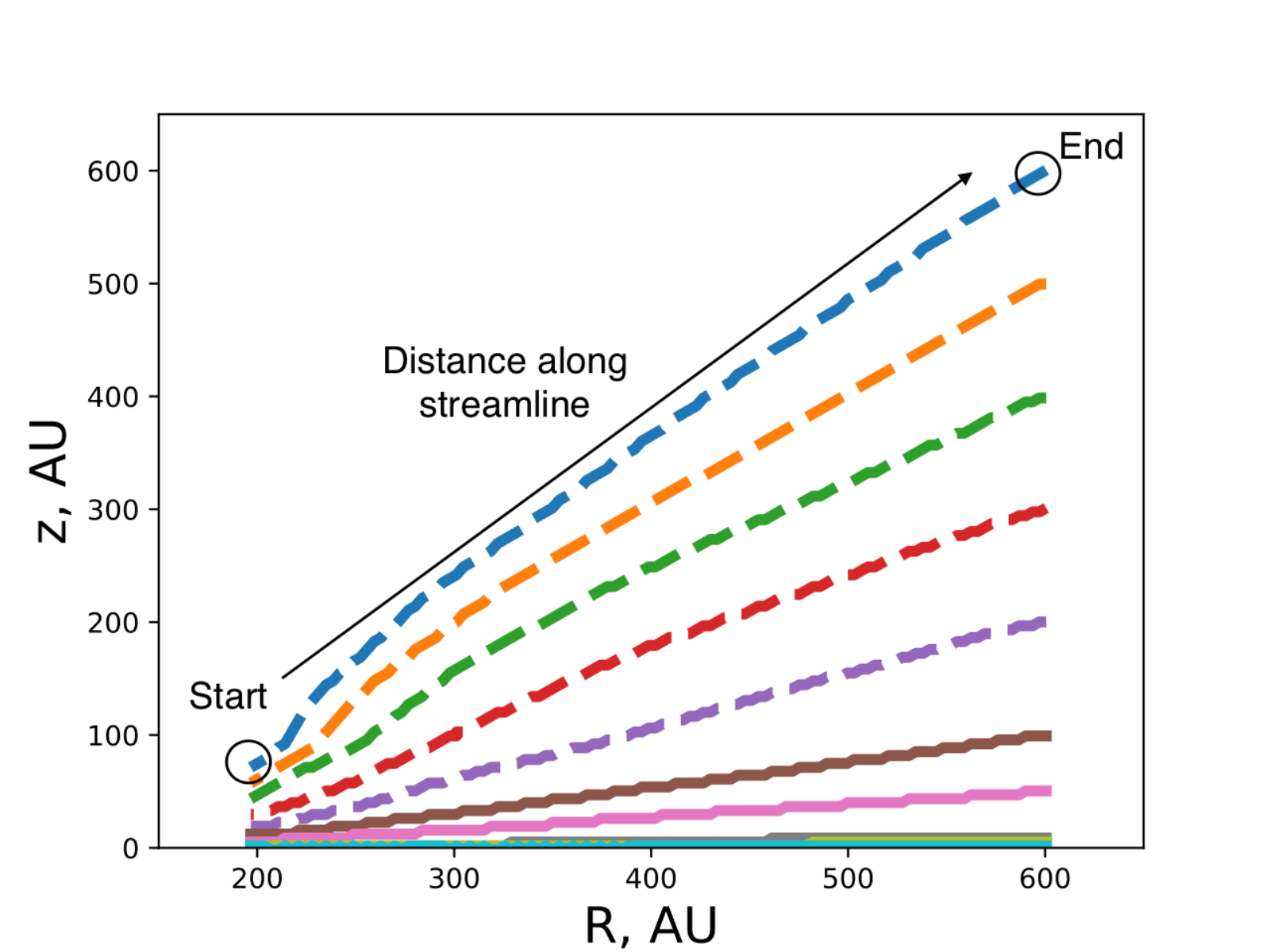}
	\includegraphics[width=9.2cm]{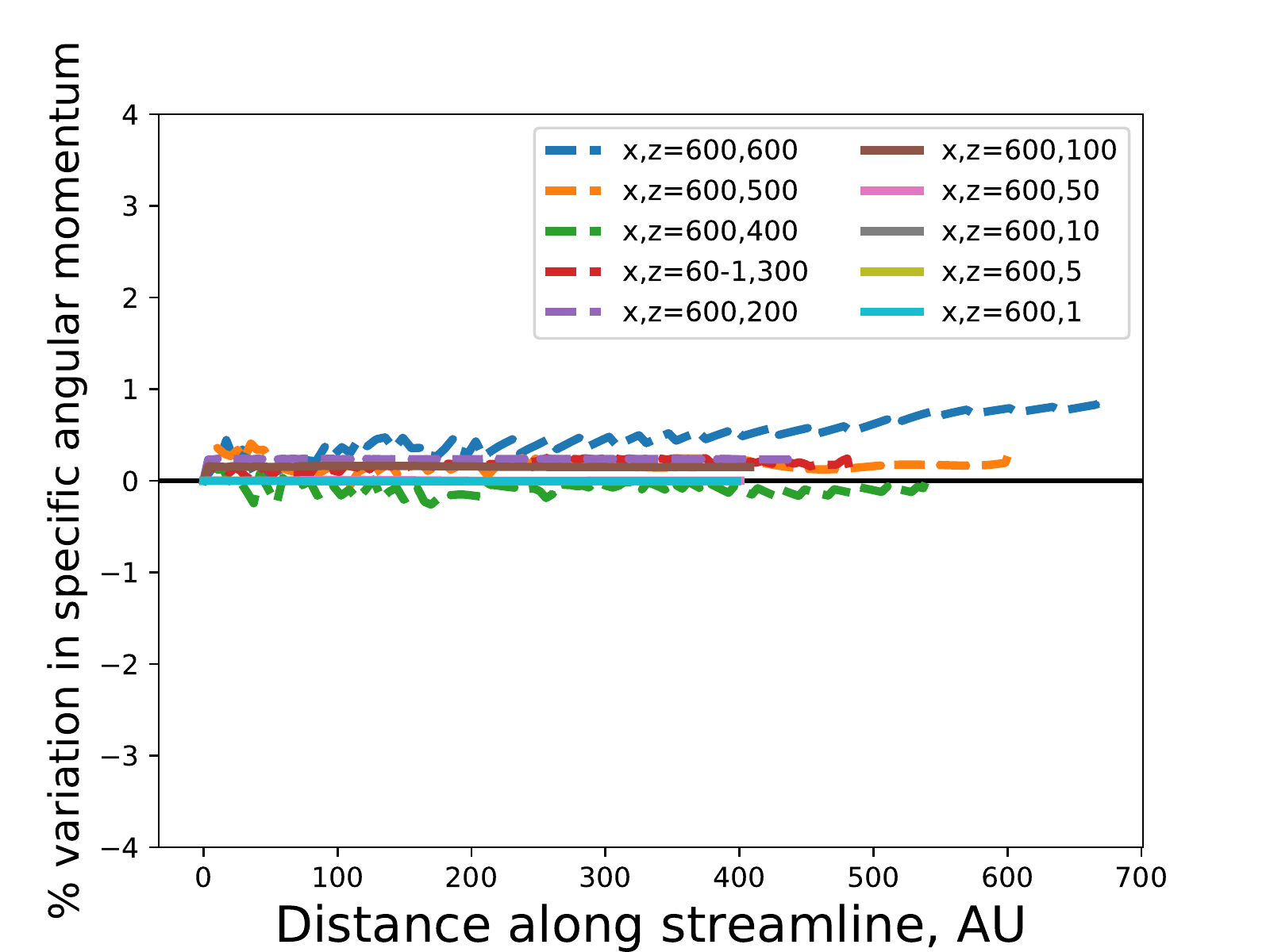}
	\includegraphics[width=9.2cm]{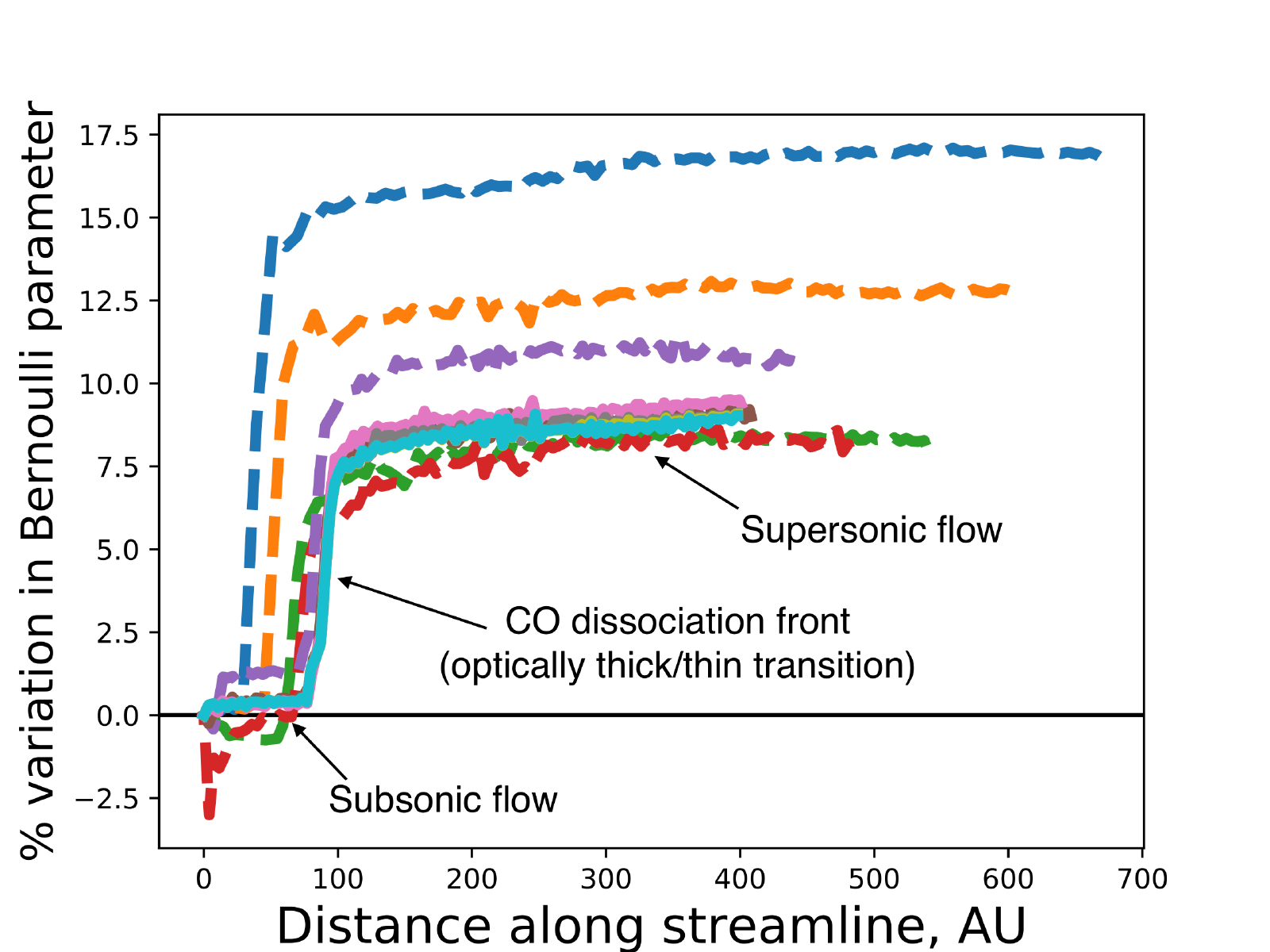}

	\caption{The upper panel shows a series of streamlines from model B (disc with a 200AU outer edge). The middle panel is the percentage variation of specific angular momentum along these streamlines from the disc edge--outwards, with the legend denoting the streamline end point in AU. The lower panel shows the percentage variation in Bernoulli parameter. The Bernoulli parameter is conserved until the steep variation in pressure at the sonic point (also the CO dissociation front) introduces
numerical errors in our evaluation of  the integral $\textrm{d}P/\rho$ in equation \ref{eqn:bernoulli}.  Beyond the pressure jump the Bernoulli parameter is again conserved.   }
	\label{fig:specAng}
\end{figure}

\subsubsection{Conservation along streamlines}
\label{sec:streamlineconvsec}
We checked the variation of specific angular momentum and Bernoulli parameter 
\begin{equation}
	\frac{v^2}{2} +\Psi + \int \frac{\textrm{d}P}{\rho} = \textrm{Constant}
	\label{eqn:bernoulli}
\end{equation}
which should be invariant along streamlines in a steady state flow. {Here $v$, $\Psi$ and $\rho$ are the local velocity, gravitational potential and density}. Note that we {evaluate the pressure term $\textrm{d}P$ } in the Bernoulli parameter in an approximate way by just taking the difference over a local length interval of the streamline. We checked \textit{a posteriori} that the flow is barotropic along a streamline (i.e. that the density is single valued function of pressure).  The specific angular momentum and Bernoulli parameter are illustrated in the case of model B in Figure \ref{fig:specAng}.  Both quantities are tightly conserved along streamlines, with deviations from perfect conservation arising from a mixture of the approximate way in which we evaluate the $\textrm{d}P$ term and small deviations from steady state. At the transition to a supersonic flow (at an optical depth $\sim1$ and, in this case, the CO dissociation front)  there is a large pressure gradient, which introduces numerical errors into our evaluation of the $\frac{\textrm{d}P}{\rho}$ term and corresponds to a jump in Bernoulli parameter. However once through to the supersonic flow the Bernoulli parameter is again conserved. We have checked that the
mass and momentum fluxes are conserved within the region of
rapidly changing thermodynamic variables. 

\begin{table*}
    \centering
    \begin{tabular}{c c c c c c c}
    \hline
    Model ID     & $\dot{M}$ 2D 64 & 2D 128 & 2D 256 & $\dot{M}$ 1D\\
   	& ($10^{-6}M_{\odot}$\,yr$^{-1}$) & ($10^{-6}M_{\odot}$\,yr$^{-1}$) & ($10^{-6}M_{\odot}$\,yr$^{-1}$) &  ($10^{-6}M_{\odot}$\,yr$^{-1}$)   \\
    \hline 
    A &  {1.9}  &  {2.2} &  {2.2} & 0.6    \\
    B & {4.8} &  {4.8}  &   {4.8}& 3.1\\
    C & {6.7} & {6.9}  & {6.8} &  5.8 \\
    \hline
    \end{tabular}
    \caption{The variation of the 2D mass loss rates with spatial resolution. The equivalent 1D mass loss rate is included for reference. Note that we evaluate the mass loss rate in the lower resolution models only where the streamlines originate from the outer half of the disc ($R>R_d/2$). This is because in the higher resolution models all of the mass loss originates from the outer half of the disc, and the inner region of the lower resolution models is poorly resolved.  }
    \label{tab:MdotResolution}
\end{table*}

\subsubsection{Convergence with spatial resolution}
There are two types of resolution element in these calculations: grid cells which quantify the spatial resolution and solid angle elements making up the \textsc{healpix} sampling of the sky from each point on the grid. Owing to the computationally expensive nature of these calculations we make a first exploration of resolution convergence with spatial resolution by coarsening our grid from $256\times256$ cells to $128\times128$ and $64\times64$ and hence using relatively cheap models. We also explore the effect of using higher  \textsc{healpix} resolution (which makes the most expensive part of the calculation and order of magnitude, or more, more expensive) in the case of the $64\times64$ cell model. 

 In our fiducial models all of the mass loss comes from the outer half of the disc  ($R>R_d/2$, see Figures \ref{fig:streamOrigins}, \ref{fig:streamOriginsScatter}). At lower resolution when the pressure gradient in the inner disc is very poorly resolved material can artificially be driven from the inner disc. Therefore, when computing the mass loss rate in the lower resolution models we only consider that from streamlines that originate from $R>R_d/2$. 

The mass loss rate at different spatial resolution is given in Table \ref{tab:MdotResolution}. In each case the variation in mass loss rate with resolution is small. Although we cannot and do not attempt to claim to have fully converged with spatial resolution (something which can be extremely challenging to do and is rarely genuinely achieved), it is promising that the mass loss rate appears to be insensitive to the spatial resolution over the range we have considered.

\subsubsection{Convergence with HEALPIX resolution}
In the case of the $64\times64$ cell spatial grid model we increased the level of healpix refinement from $l=0$ (the fiducial value here) to $l=1$ and $l=2$, where the number of \textsc{healpix} zones is $N=12\times4^l$. That is we consider 12, 48 and 192 rays per cell (the $l=2$ case is roughly an order of magnitude more expensive than the $l=0$ case, which is why we explore convergence with \textsc{healpix} rays on a coarse spatial grid). The mass loss rate for different levels of \textsc{healpix} refinement in the case of model A are given in Table \ref{tab:hlevVar}. As with spatial resolution, we cannot claim convergence, but there is no strong variation in the mass loss rate with \textsc{healpix} refinement. The reason for the slight increase in mass loss rate that can happen at higher \textsc{healpix} resolution is that there is slightly higher temperature/pressure and hence driving of a wind from the disc surface layers.

\begin{table}
    \centering
    \begin{tabular}{c c c }
    \hline
    healpix level    &Rays per cell &  $\dot{M}$ \\
   	& & ($10^{-6}M_{\odot}$\,yr$^{-1}$) \\
    \hline 
    0 & 12& {1.9}  \\
    1 & 48&  2.1  \\
    2 & 192& {2.0}\\    
    \hline
    \end{tabular}
    \caption{Mass loss rates in a low resolution ($64\times64$ cell) version of model A, varying the level of healpix refinement (number of rays used to sample $4\pi$ steradians per cell). }
    \label{tab:hlevVar}
\end{table}

\subsection{Chemical composition}
\label{sec:chemistry}
\begin{figure*}
	\hspace{0cm}
	\includegraphics[width=2.2cm]{modelA.pdf}

	\vspace{0.2cm}
	\hspace{-0.5cm}
	\includegraphics[width=5.44cm]{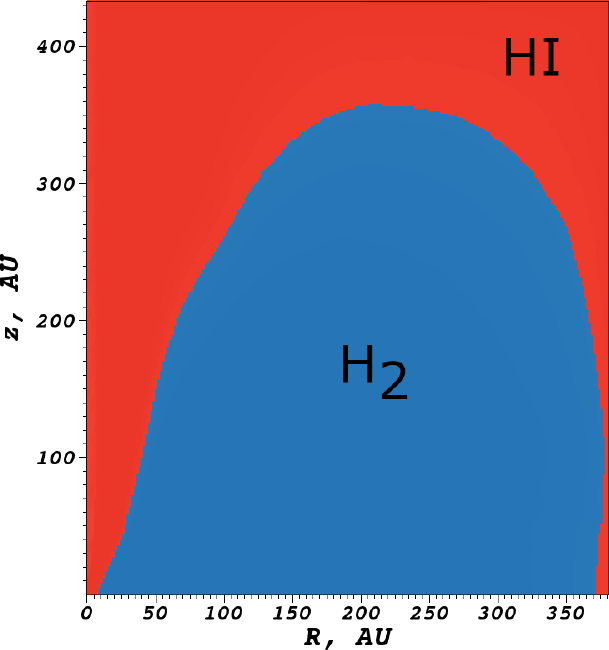}
	\includegraphics[width=5.44cm]{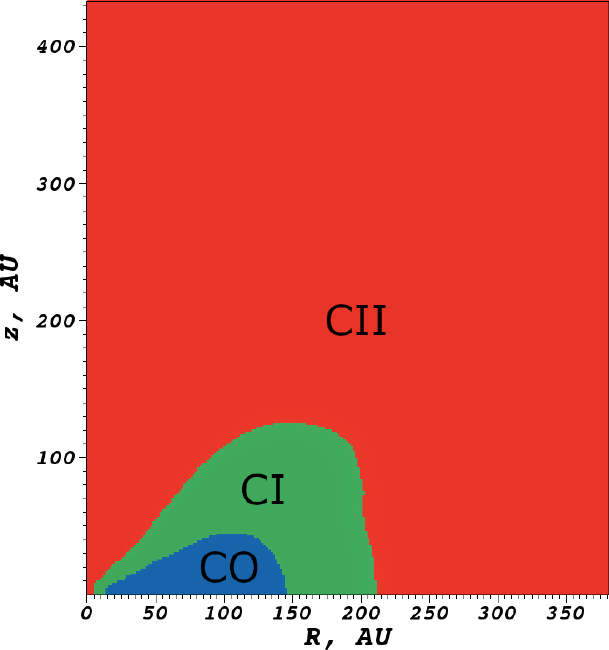}
	\includegraphics[width=6.78cm]{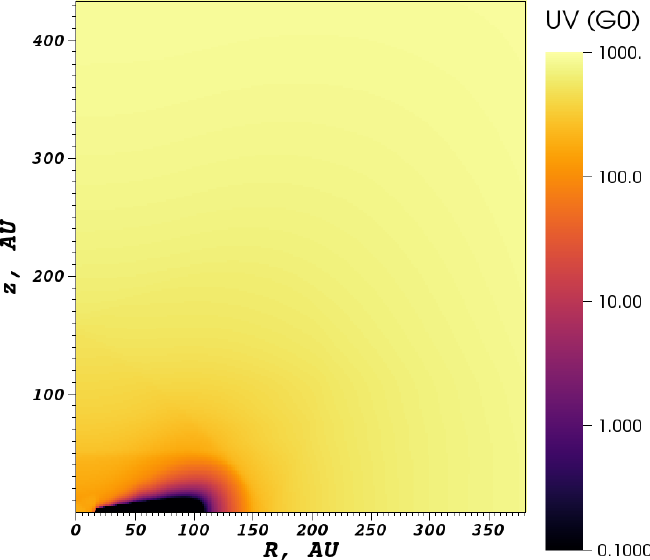}	

	\vspace{0.4cm}	
	\hspace{0cm}
	\includegraphics[width=2.2cm]{modelB.pdf}
	
	\hspace{-0.5cm}
	\includegraphics[width=5.44cm]{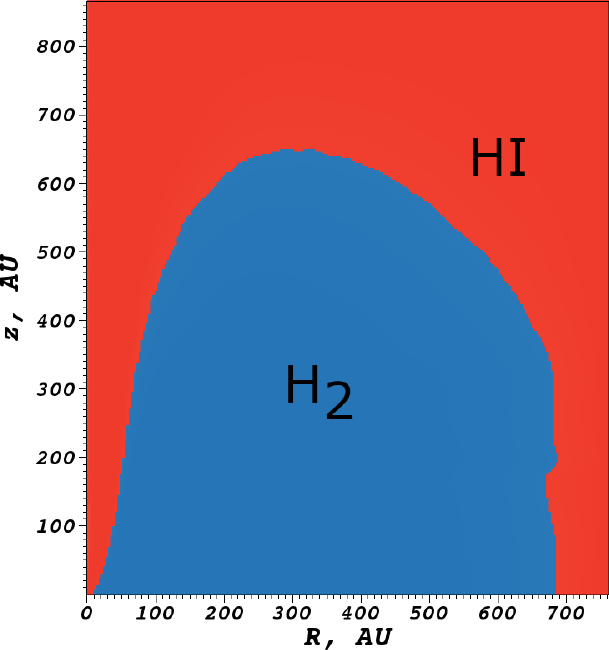}
	\includegraphics[width=5.44cm]{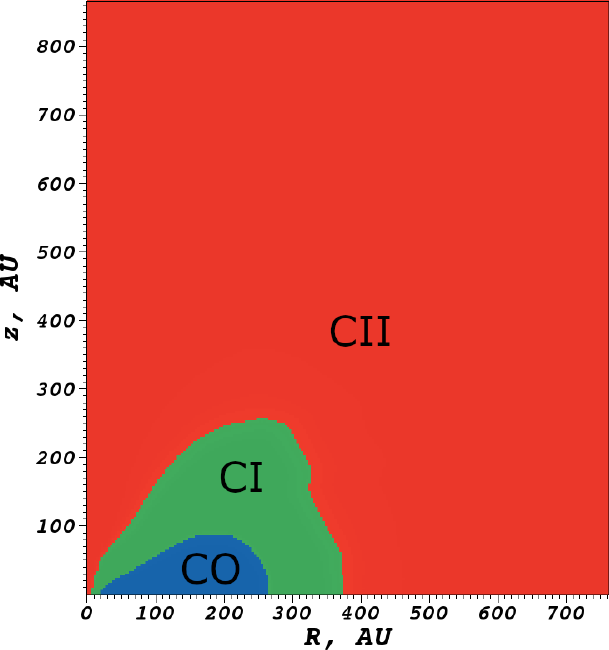}
	\includegraphics[width=6.78cm]{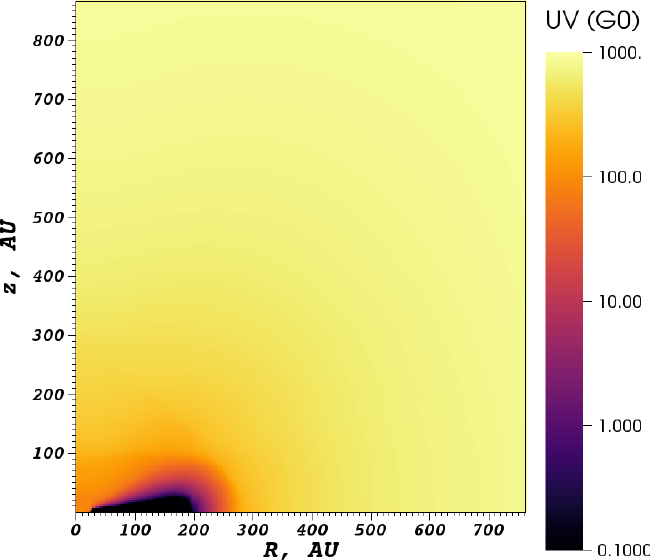}

	\vspace{0.4cm}	
	\hspace{0cm}
	\includegraphics[width=2.2cm]{modelC.pdf}
		
	\hspace{-0.5cm}
	\includegraphics[width=5.44cm]{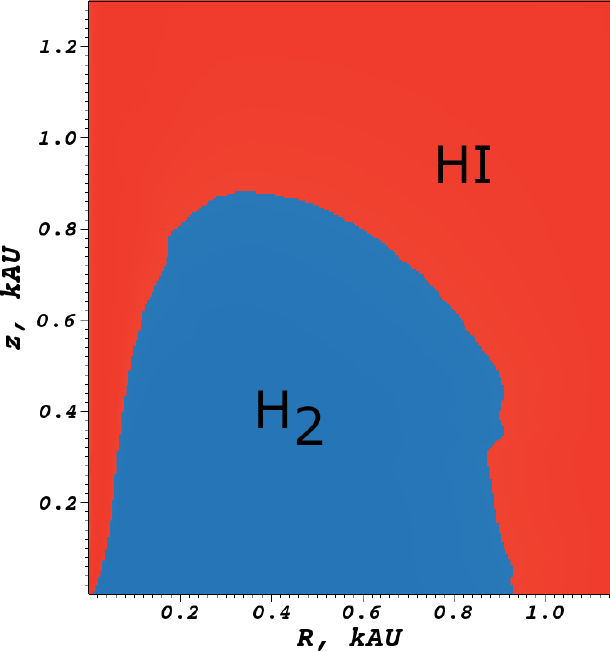}
	\includegraphics[width=5.44cm]{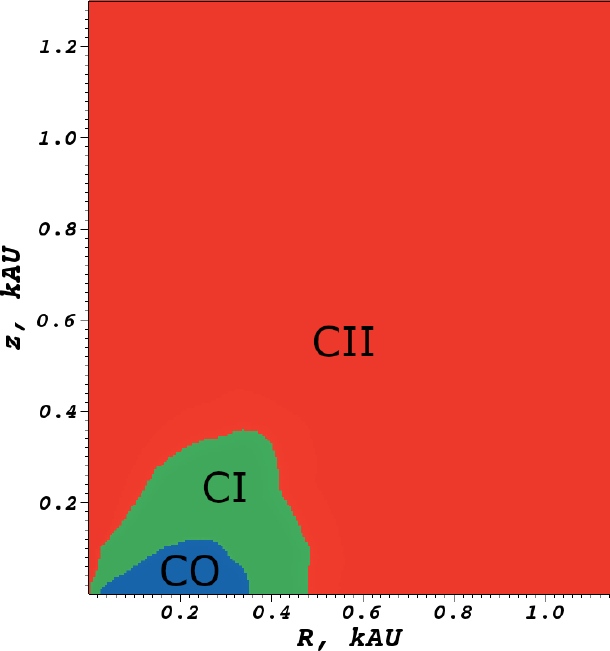}
	\includegraphics[width=6.78cm]{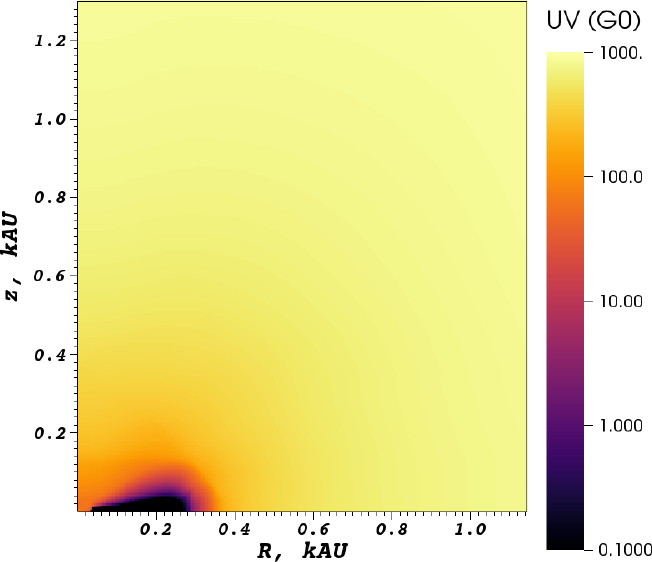}

	\caption{{Maps of some chemical properties and the UV field of our models. The left hand column compares the abundances of atomic and molecular hydrogen, denoting molecular hydrogen dominated gas as blue and atomic hydrogen dominated zones as red.  The middle column illustrates in which regions CO, C\,I and C\,II dominate in abundance over one another, with CO gas in blue, C\,I in green and C\,II in red. The right column is the UV field (in Habing units, G$_0$). Models are A--C from top to bottom.}}
	\label{fig:Chem}
\end{figure*}

To solve for the dynamics our calculations necessarily solve a PDR chemistry network and therefore naturally yield the chemical structure of the disc/wind. Figure \ref{fig:Chem} illustrates some features of the composition of our models. The right hand panels are maps of the UV field strength, which attenuates sharply near the halo boundary and rapidly drops to a negligibly low value within the disc itself. Ionisation and dissociation contributing to the chemistry of the disc itself is therefore more sensitive to the cosmic ray ionisation rate than the ambient UV field \citep[for more information on disc chemistry including assessments of the cosmic ray contribution see e.g.][]{2014ApJ...794..123C, 2015ApJ...799..204C, 2019MNRAS.484.1563W}. Though as we will shortly see parts of the wind are molecular, so may be identified as part of the disc in molecular line observations and certainly would be influenced by the ambient UV field. 

The left and middle columns of of Figure \ref{fig:Chem} illustrate in which zones certain species dominate in abundance over one another. In the left panels the blue regions are molecular hydrogen and the red atomic and so the H--H$_2$ transition surface is at their interface. Similarly in the middle hand panels we illustrate which of CO, C\,I and C\,II dominate over one another in certain zones. The CO halo region extends beyond our imposed disc structure. This is then enveloped by a layer of C\,I, beyond which the medium becomes warmer and C\,II is most abundant of these 3 species. {Comparing the chemical distribution with the dynamical properties of Figures \ref{fig:rhoVT} and \ref{fig:mach}, the carbon cycle is more strongly correlated with the dynamics than the H--H$_2$ transition.}

\begin{figure}
	\hspace{2.7cm}
	\includegraphics[width=2.2cm]{modelA.pdf}
	
	\includegraphics[width=8.55cm]{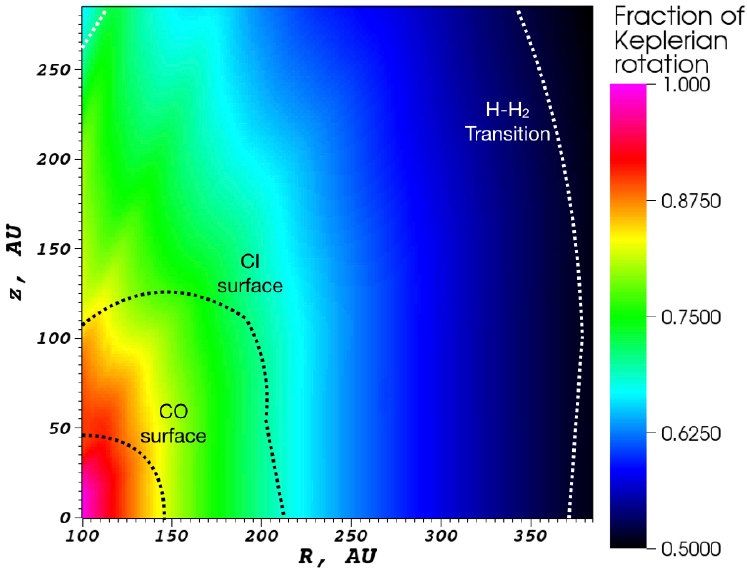}	

	\vspace{0.25cm}
	\hspace{2.7cm}
	\includegraphics[width=2.2cm]{modelB.pdf}
	
	\includegraphics[width=8.55cm]{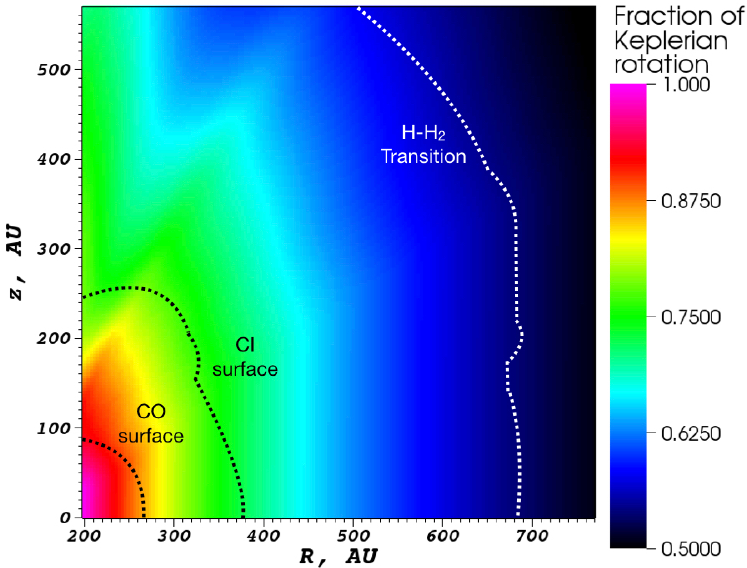}	

	\vspace{0.25cm}	
	\hspace{2.7cm}
	\includegraphics[width=2.2cm]{modelC.pdf}	

	\includegraphics[width=8.55cm]{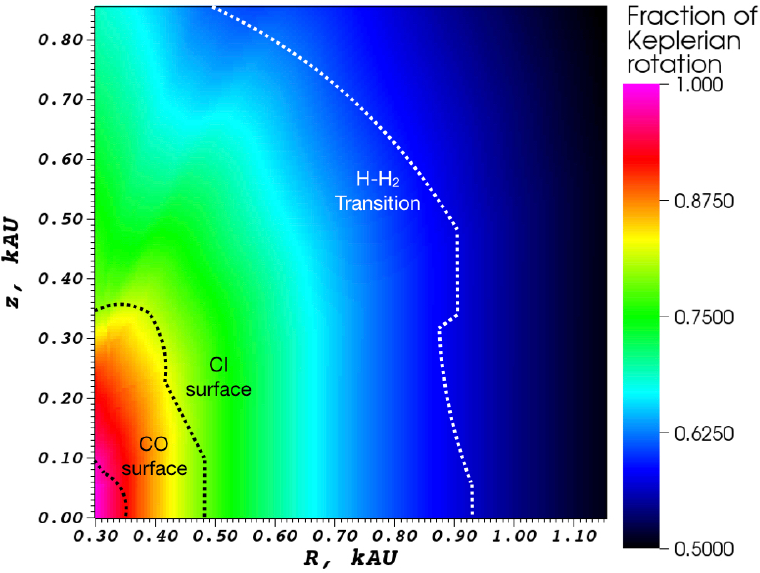}

	\caption{The ratio of azimuthal to Keplerian velocity in the winds of our models {(i.e. a value of unity is Keplerian)}. {The disc outer edge is at the left hand edge of each panel.} The dashed lines denote the surfaces at which CO becomes less abundant than C\,I (the CO surface) and at which C\,I becomes less abundant than C\,II (the C\,I surface).  Sub-Keplerian rotation provides a possible means of detecting externally driven winds. }
	\label{fig:vPhiOnVkep}
\end{figure}

One of the predictions of 1D models is the the azimuthal velocity in the flow is sub-Keplerian \citep{2016MNRAS.457.3593F, 2016MNRAS.463.3616H} which might be identifiable observationally. Indeed observational hints of sub-Keplerian rotation were found in the case of the disc IM Lup by \cite{2018A&A...609A..47P}. Figure \ref{fig:vPhiOnVkep} shows the spatial distribution of $v_\phi/v_{\textrm{kep}}$ for each model. Marked on with dashed lines are the surfaces at which C\,I transitions to being dominant over CO (the CO surface) and where C\,II transitions to being dominant over C\,I (the C\,I surface).  Deviations from Keplerian rotation in the wind can be very strong, but in the models considered here this is mainly in the atomic and ionised carbon zones. Atomic carbon may offer a promising alternative for probing environmentally driven winds and detecting sub-Keplerian rotation, though we will explore observational signatures using synthetic observations in a subsequent paper.

\begin{figure}
	\includegraphics[width=9.4cm]{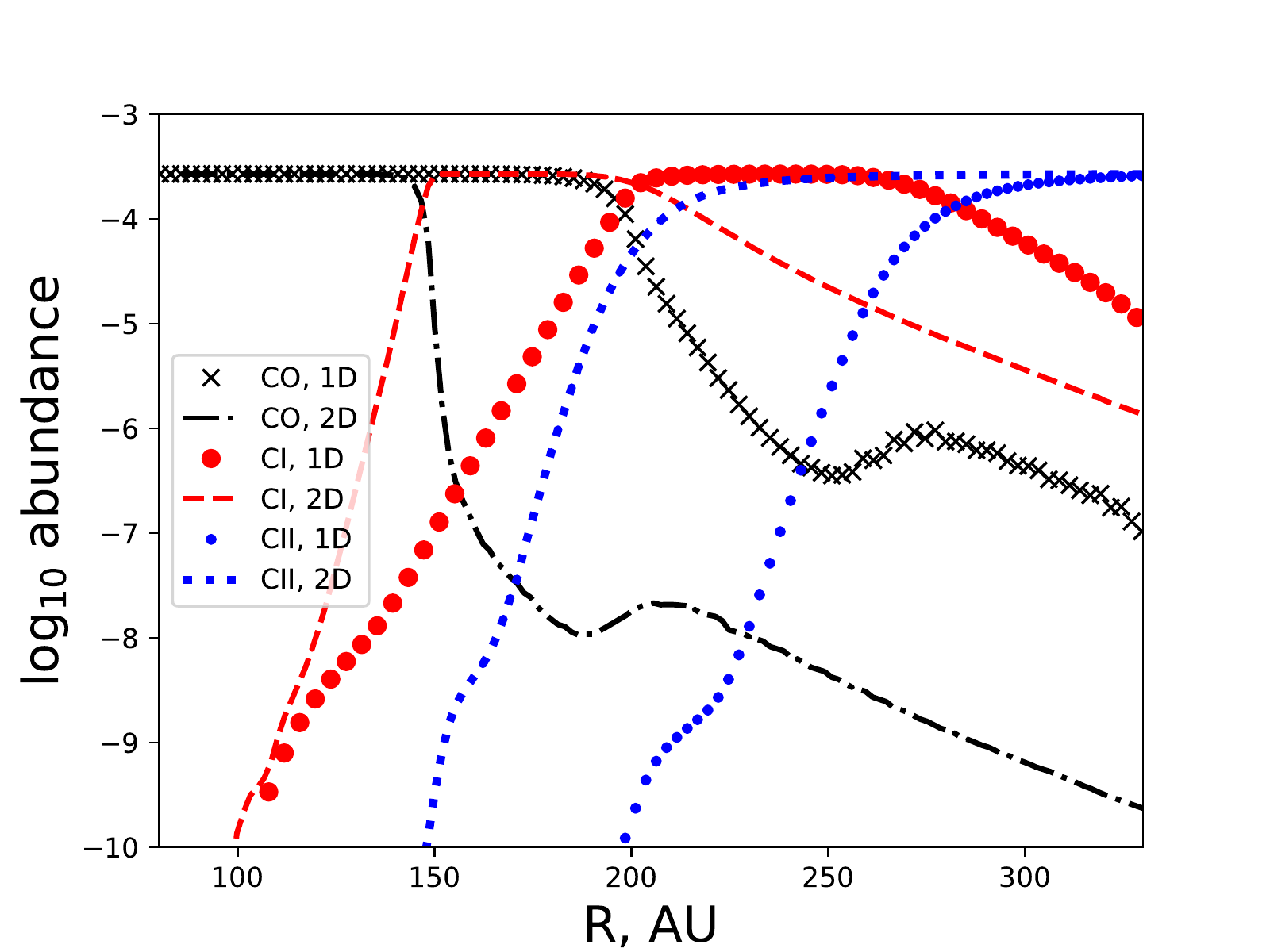}

	\vspace{0.25cm}	
	\includegraphics[width=9.4cm]{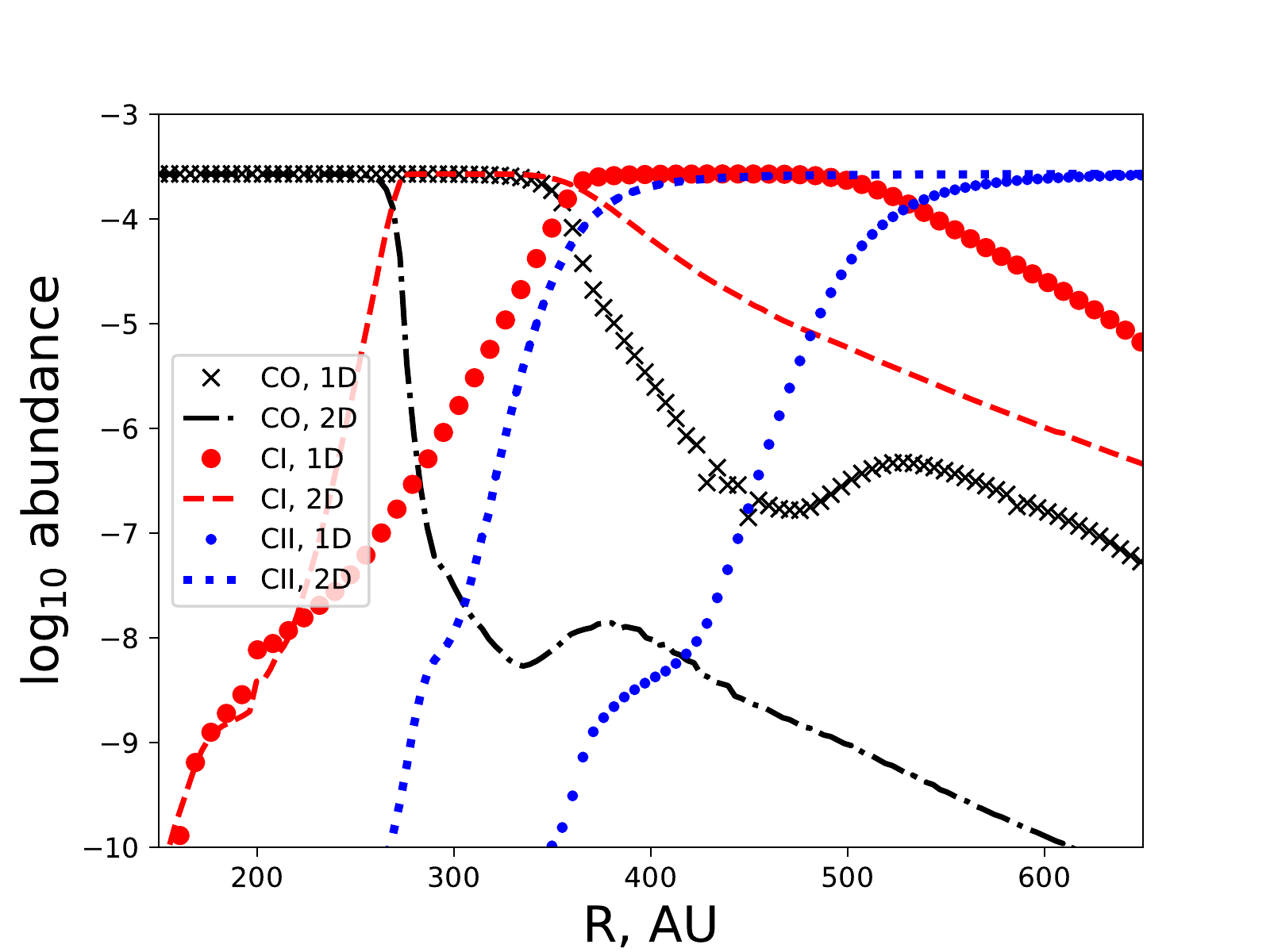}	

	\vspace{0.25cm}
	\includegraphics[width=9.4cm]{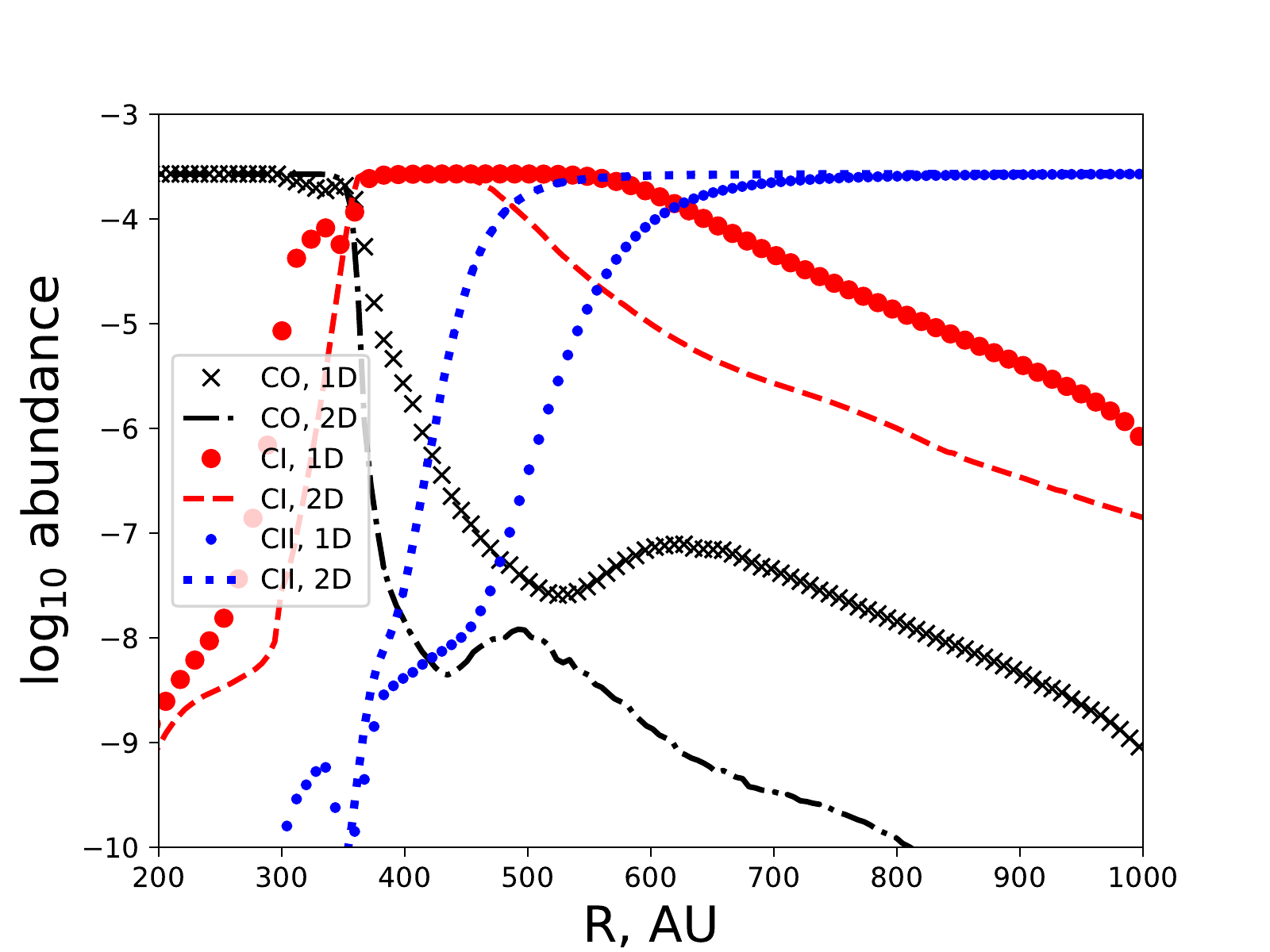}
	\caption{The mid-plane chemical abundance profile of some key coolant species. The solid lines are the 2D models and points from the 1D models.  Panels are models A-C from top to bottom. }
	\label{fig:midplane_chem}
\end{figure}

Finally, in Figure \ref{fig:midplane_chem}  we compare the chemical profile of some species in the 1D models with the mid-plane chemical profile from our 2D models. The panels show models A-C from top to bottom. Solid lines are the 2D distribution and points the 1D. There is a common theme, at least in these models, which is that the radial extent of molecular gas is significantly reduced in the 2D models compared to 1D. This is because the 1D models assume that exciting radiation only propagates inwards radially and that other trajectories are infinitely optically thick. Heating of the mid-plane from trajectories above the mid-plane is responsible for dissociating gas more effectively than the 1D models predict, decreasing the CO extent. The consequences of this are that (at least) 2D models will be required to properly predict observable signatures of externally driven winds, and also that they will likely be harder to detect than we might have expected from prior 1D models, at least at $10^3$\,G$_0$ UV field strengths.

\section{Summary and conclusions}
We have computed the first 2D PDR-dynamical models of externally FUV irradiated protoplanetary discs, including direct computation of the PDR microphysics and line cooling in 3D. Previously, only 1D (usually semi-analytic) calculations of this have been possible. In this paper we aim to make a first study of the multidimensional flow structure, calculate mass loss rates and compare with the 1D calculations that are important for providing mass loss rates for disc viscous evolutionary models. We draw the following main conclusions from this work. \\

\noindent 1) Mass loss rates from our 2D models are slightly higher than in equivalent 1D cases, to within a factor $\sim4$. Differences can arise from a few factors. The 1D model mass loss rates are enhanced relative to the 2D because they consider the mid-plane only and assume that this densest part of the flow represents what would propagate spherically outwards from an entire scale height at the disc outer edge. Nevertheless, this is more than compensated for by the fact that we find that only about half of the mass loss comes from the very disc outer edge (it is all assumed to originate from there in the 1D models) with the rest coming from the disc surface.  We conclude that 1D estimates of mass loss rates can reasonably be used in other applications, especially since they are expected to be conservative estimates. This is an important result since 1D evolutionary models find that external photoevaporation is important for disc mass and radius evolution. It thus controls the viscous timescale and hence the surface density evolution of the entire disc and so could even affect inner planet formation.   \\

\noindent 2) Aside from the mass loss, a key difference between the 1D and 2D models is the chemical structure. This arises because the 1D calculations assume that exciting radiation only propagates radially inwards and that other trajectories (i.e. from above the mid-plane) are infinitely optically thick. This means that CO is dissociated at smaller radii in 2D models compared to 1D. Although 1D mass loss rates are reasonable (and conservative) multidimensional models such as those in this paper are  going to be essential for predicting observables of externally evaporating discs using synthetic observations. \\

\noindent 3) Sub-Keplerian rotation in the wind was a signature of external photoevaporation predicted by 1D models \citep{2016MNRAS.457.3593F, 2016MNRAS.463.3616H}. We do still anticipate that sub-Keplerian rotation could be a viable signature of external photoevaporation; however at the intermediate UV field strengths considered here it is anticipated to be difficult to detect in CO (at lower UV field strengths we anticipate that stronger sub-Keplerian deviations will be in CO-bearing parts of the flow). Rather, at intermediate UV field strengths our models suggest that atomic carbon lines could offer a promising alternative tracer of the wind and sub-Keplerian rotation, which we will explore with subsequent synthetic observations.  


\section*{Acknowledgements}

We thank the referee for their swift and constructive review that improved the manuscript. We {thank} James Owen for useful discussions and his comments on this manuscript. 
TJH is funded by an Imperial College London Junior Research Fellowship. CJC  has been supported by the DISCSIM project, grant agreement
341137 funded by the European Research Council under ERC-
2013-ADG. This work was performed using the DiRAC Data Intensive service at Leicester,
operated by the University of Leicester IT Services, which forms part of the
STFC DiRAC HPC Facility (www.dirac.ac.uk). The equipment was funded by BEIS
capital funding via STFC capital grants ST/K000373/1 and ST/R002363/1 and STFC
DiRAC Operations grant ST/R001014/1. DiRAC is part of the National
e-Infrastructure.
The work also used the COSMOS
Shared Memory system at DAMTP, University of Cambridge operated
on behalf of the STFC DiRAC HPC Facility. This equipment is
funded by BIS National E-infrastructure capital grant ST/J005673/1
and STFC grants ST/H008586/1, ST/K00333X/1. DiRAC is part of
the National E-Infrastructure. It also used the DiRAC Data Analytics system at the
University of Cambridge, operated by the University of Cambridge
High Performance Computing Serve on behalf of the STFC DiRAC
HPC Facility (www.dirac.ac.uk). This equipment was funded
by BIS National E-infrastructure capital grant (ST/K001590/1),
STFC capital grants ST/H008861/1 and ST/H00887X/1, and STFC
DiRAC Operations grant ST/K00333X/1. The main DiRAC project ID for these calculations was dp100 (PI: J. E. Owen). 

\bibliographystyle{mnras}
\bibliography{molecular}

\label{lastpage}

\end{document}